\documentclass[journal]{IEEEtran}

\usepackage{cite}
\usepackage{array}
\usepackage{url}
\usepackage[dvips]{epsfig}
\usepackage[latin1]{inputenc}
\usepackage[T1]{fontenc}
\usepackage{url}
\usepackage{booktabs}
\usepackage{amsmath}

\begin{document}

\title{Artificial Intelligence Based Cognitive Routing for Cognitive Radio Networks}

\author{Junaid Qadir

\thanks{This work has been supported by Higher Education Commission (HEC), Pakistan under the NRPU programme.}
\thanks{Junaid Qadir (\emph{junaid.qadir@seecs.edu.pk}) is with the Electrical Engineering Department at the School of Electrical Engineering and Computer Science (SEECS) at the National University of Sciences and Technology (NUST), Pakistan.}}

\maketitle

\begin{abstract}

Cognitive radio networks (CRNs) are networks of nodes equipped with cognitive radios that can optimize performance by adapting to network conditions. While cognitive radio networks (CRN) are envisioned as intelligent \emph{networks}, relatively little research has focused on the network level functionality of CRNs. Although various routing protocols, incorporating varying degrees of adaptiveness, have been proposed for CRNs, it is imperative for the long term success of CRNs that the design of \emph{cognitive routing protocols} be pursued by the research community. Cognitive routing protocols are envisioned as routing protocols that fully and seamless incorporate AI-based techniques into their design. In this paper, we provide a self-contained tutorial on various AI and machine-learning techniques that have been, or can be, used for developing cognitive routing protocols. We also survey the application of various classes of AI techniques to CRNs in general, and to the problem of routing in particular. We discuss various decision making techniques and learning techniques from AI and document their current and potential applications to the problem of routing in CRNs. We also highlight the various inference, reasoning, modeling, and learning sub tasks that a cognitive routing protocol must solve. Finally, open research issues and future directions of work are identified.

\end{abstract}

\section{\uppercase{Introduction}}
\label{sec:introd}

In cognitive radio networks (CRNs), nodes are equipped with \emph{cognitive radios} (CRs) that can sense, learn, and react to changes in network conditions. Mitola envisioned that CRs could be realized through incorporation of substantial computational or artificial intelligence (AI)---particularly, machine learning, knowledge reasoning and natural language processing \cite{mitola2006cognitive}---into SDR hardware. In a modern setting, this is achieved by incorporation of a cognitive engine (CE) using various AI-based techniques through which the CR adapts to the network conditions to satisfy some notion of optimality \cite{he2010survey}. CRs have also been proposed for a wide range of applications including intelligent transport systems, public safety systems, femtocells, cooperative networks, dynamic spectrum access, and smart grid communications  \cite{he2010survey}\cite{akyildiz2006next}. CR promises to dramatically improve spectrum access, capacity, and link performance while also incorporating the needs and the context of the user \cite{he2010survey}. CRs are increasingly being viewed as an essential component of next-generation wireless networks \cite{akyildiz2006next} \cite{haykin2005cognitive}.

Although cognitive behavior of CRNs can enable diverse applications, perhaps the most cited application of CRNs is dynamic spectrum access (DSA)\footnote{DSA is such a dominantly cited application of CRNs that DSA and CRN are often assumed to be synonymous incorrectly. CRNs, in fact, is a much broader concept allowing for diverse applications representing intelligent behavior \cite{fette2009cognitive}.} \cite{fette2009cognitive}. DSA is proposed as a solution to the problem of \emph{artificial spectrum scarcity} that results from static allocation of available wireless spectrum using the command-and-control licensing approach \cite{fette2009cognitive}. Under this approach, licensed applications represented by \emph{primary users} (PUs) are allocated exclusive access to portions of the available wireless spectrum prohibiting other users from access even when the spectrum is idle. With most of the radio spectrum already being licensed in this fashion, innovation in wireless technology is constrained. The problem is compounded by the observation, replicated in numerous measurement based studies world over, that the licensed spectrum is grossly underutilized \cite{akyildiz2006next} \cite{fette2009cognitive}. The DSA paradigm proposes to allow secondary users (SUs), also called cognitive users, access to the licensed spectrum subject to the condition that SUs do not interfere with the operations of the primary network of incumbents.

While CRs have been defined differently \cite{he2010survey}, the following tasks are considered integral to them:
\emph{i)} \emph{observation} or \emph{awareness}, \emph{ii)} \emph{reconfiguration}, and \emph{iii)} \emph{cognition}. In this paper, we will be occupied mostly with cognition as we seek to build cognitive, AI-based, routing protocols. Cognition subsumes both reasoning and learning with \emph{reasoning} being the process of finding the appropriate action for particular situations to meet some system target, and \emph{learning} being the process of accumulating knowledge based on the results of previous actions \cite{he2010survey} \cite{gavrilovskalearning}. Generally speaking, cognition for a CR entails understanding and reasoning about the radio environment so that informed decisions may be taken to optimize the performance of the radio and of the overall network.

Both learning and reasoning are essential elements of cognition and a lot of research attention has rightly focused on incorporating cognition in CRs. However, while incorporating learning and adaptiveness into CRs is highly desirable, the vision of a `\emph{cognitive network}' will not be realized until the networks, and the network layer functions, seamlessly incorporate intelligence. Cognitive networks are envisioned as intelligent networks that perceive current conditions to plan, decide and act while catering to the network's overall end-to-end goals \cite{thomas2007cognitive} \cite{fortuna2009trends}. Cognitive networking broadly encompasses models of cognition and learning that have been defined for CRs but are distinguished from isolated CRs in its emphasis on its networking wide and end-to-end scope. In previous work on cognitive networks, Mahonen et al. proposed a cognitive resource manager as a framework for \emph{network-wide} optimization of radio resources, and proposed utilizing machine-learning techniques to manage cross-layer optimization \cite{mahonen2006cognitive} \cite{mahonen2004cognitive}. Some ten years ago, Clark et al. proposed that Internet must have a \emph{knowledge plane} distinct from the data and the control planes that will allow building up an intelligent network capable of setting itself up given high level instructions, adapt itself to changing requirements, manage itself to automatically discover anomalies, and automatically fix problems or explain why it cannot do so \cite{clark2003knowledge}. Clark et al. noted that building such a `cognitive network' would require AI-based cognitive techniques and not just incremental algorithmic techniques.

%While Thomas et al. include wired networks in their conception of cognitive networks \cite{thomas2007cognitive}, our work focuses exclusively on wireless cognitive networks or alternatively \emph{``cognitive radio networks''}.

To help CRNs become \emph{cognitive networks}, it is imperative that intelligence be integrated into the fabric of CRN architecture and protocols across the stack. Some challenges that confront learning algorithms in CRNs, as identified in \cite{bkassinysurvey}, are as follows:
\begin{enumerate}
\item Learning algorithms have to operate in certain cases in unknown RF environments without any supervision.
\item Learning algorithms have to operate in environments that are only partially observable.
\item Learning algorithms for CRNs require distributed algorithms due to the decentralized nature of CRNs and are properly envisioned in multi-agent learning which are more challenging that single-agent learning scenario.
\end{enumerate}

\vspace{2mm}
\emph{Contributions of this paper:} In this paper, we weave together ideas from multiple disciplines (such as optimization theory, game theory, machine learning, artificial intelligence, control theory, and economics) to present a cogent and holistic overview of techniques that can be useful for network-layer decision making in CRNs. This task has been non-trivial due to the multi-disciplinary nature of CRN research which is compounded by the fact that many of parent fields use different terminology and notation for similar concerns. Previous survey articles that are similar to this work have focused mainly on application of machine-learning and AI techniques to problems of spectrum sensing, power control, and adaptive modulation in CRNs \cite{he2010survey} \cite{bkassinysurvey}. To the best of our knowledge, this is the first survey article that focuses on the application of AI techniques to the problems of modeling, design and analysis of network-layer issues (in particular, the problems of routing and forwarding) in CRNs.

In this paper, the basic concepts of relevant AI techniques are presented and their applications to CRNs, particularly for routing, are highlighted. While this paper attempts to be self-contained, it is not intended as a exhaustive document keeping in view the breadth of topics covered. It has been attempted to provide links to more comprehensive resources on specialized topics where ever appropriate.

\vspace{2mm}
\emph{Organization of this paper:} The rest of the paper is organized as follows. Section \ref{sec:background} presents the necessary machine-learning background before we discuss \emph{decision and planning techniques} in section \ref{sec:decandplanning}, and \emph{learning techniques} in section \ref{sec:learning}, respectively. A survey of existing routing protocols for CRNs is presented in section \ref{sec:routing} and it is shown that while these protocols do support certain adaptive features, more work needs to be done to build AI-enabled cognitive routing protocols for CRNs. Some important tasks that an AI-enabled cognitive routing protocol must implement are discussed in section \ref{sec:cogroutingtasks}. Open research issues and future research directions are identified in section \ref{sec:openissues}. Finally, the paper is concluded in section \ref{sec:concl}.

\section{\uppercase{Background: Machine Learning}}
\label{sec:background}

For a radio to be deemed a \emph{cognitive} radio, it is necessary for it to be equipped with the ability of learning \cite{haykin2005cognitive}. On receiving certain environmental input, systems (e.g., animals, automata, and in our case, cognitive radios) exhibit some kind of behavior. If the system changes changing its behavior over time in order to improve its performance at a certain task, it is said to \emph{learn} from its interaction with its environment. This implies that these systems may respond differently to the same input later on than they did earlier. The field of machine learning focuses on the theory, properties and performance of \emph{learning algorithms}.

\emph{Machine learning} is a field of research that formally studies learning systems and algorithms. It is a highly interdisciplinary field building upon ideas from diverse fields such as statistics, artificial intelligence, cognitive science, information theory, optimization theory, optimal control, operations research, and many other disciplines of science, engineering and mathematics \cite{abu2012learning} \cite{russell1995artificial} \cite{mitchell1997machine} \cite{mitchell2006discipline}. Russell and Norvig \cite{russell1995artificial} describe machine learning to be the ability to ``adapt to new circumstances and to detect and extrapolate patterns''. Machine learning techniques have proven themselves to be of great practical utility in diverse domains such as pattern recognition, robotics, natural language processing, autonomous control systems. They are particularly useful in domains, like CRNs, where the agents must dynamically adapt to changing conditions.

\vspace{2mm}
\emph{Type of machine learning algorithms:} Machine learning concerns itself with a learner using a set of observations to uncover the underlying process \cite{abu2012learning}. There are principally three variations to this broad definition and machine learning can be classified into three broad classes with respect to the sort of feedback that the learner can access: \emph{i)} supervised learning, \emph{ii)} unsupervised learning, and \emph{iii)} reinforcement learning. Briefly, supervised learning is one extreme in which the learner is provided with labeled examples by its environment (alternatively, a supervisor or teacher) in a training phase through which the learner attempts to generalize so that it can respond correctly to inputs it has not seen yet. We can think of learning a simple categorization task as supervised learning. Unsupervised learning is the other extreme in which the learner receives no feedback from the environment at all. The learner's task is to organize or categorize the inputs in clusters, categories, or with reduced set of dimensions. A third alternative, closer to supervised learning than to unsupervised learning, is reinforcement learning in which although the learner is not provided feedback about what exactly the correct response should have been, it gets indirect feedback about the appropriateness of the response through a reward (or reinforcement). Reinforcement learning, therefore, depends more on exploration through trial-and-error. We will be covering these three kinds of learning in more detail later in sections \ref{supervised}, \ref{unsupervised}, and \ref{reinforcement}, respectively.

\vspace{2mm}
\emph{Previous work on applying machine learning to CRNs:} Bkassiny et al. provide a comprehensive survey of applications of machine-learning techniques in CRNs \cite{bkassinysurvey}, and divide learning applications for CRNs into two broad categories of  \emph{feature classification} and \emph{decision making}. Feature classification mainly has applications in spectrum sensing and signal classification. Decision making has diverse applications in CRNs including adaptive modulation, power control, routing and transport-layer applications \cite{bkassinysurvey}. Decision making problems can be further classified into policy making and decision rules problems. In a policy making problem, an agent determines an optimal \emph{policy} (or an optimal \emph{strategy} in game theory terminology) to determine what actions it should perform over a certain time duration. In a decision rule problem, on the other hand, the problem is formulated as hypothesis testing problem and the aim is to directly learn the optimal values of certain design and operation parameters \cite{bkassinysurvey}. Bkassiny et al. also establish the relationship between learning and optimization and show that many learning algorithms converge towards the \emph{optimal} solution concept in their respective applications (whenever it exists). Applications of machine learning to CRNs are vast \cite{clancy2007applications}  \cite{rondeau2007application}, and we shall develop a more complete picture gradually as we proceed in this paper. Interested readers are referred to the surveys \cite{he2010survey}\cite{bkassinysurvey}, and the references therein, for a comprehensive complementary treatment of general applications of machine learning to CRNs.

\subsection{\uppercase{Supervised learning}}
\label{supervised}

In supervised learning, algorithms are developed to learn and extract knowledge from a set of training data which is composed of inputs and corresponding outputs assumed to be labelled correctly by a `teacher' or a `supervisor'. To understand supervised learning, imagine a machine that experiences a series of inputs: $x_1$, $x_2$, $x_3$, and so on. The machine is also given the corresponding desired outputs $y_1$, $y_2$, $y_3$, and so on, and the goal is to learn the general function $f(\textbf{x})$ through which correct output can be determined given a new input $x_i$ (not necessarily seen in the training examples provided).

The output can be a continuous value for a regression problem, or can be a discrete value for a classification problem. The objective of supervised learning is to predict the output given any valid input. In other words, the task in supervised learning is to discover the function through which an input is transformed into output. This contrasts with `unsupervised learning' in which the example of objects are available in an unlabelled or unclassified fashion.

\vspace{2mm}
\emph{Types of supervised learning problems}: There are essentially two types of supervised learning problems---classification and regression (or estimation). Classifiers itself can be further \emph{classified} into \emph{computational classifiers} such as support vector machines (SVM), \emph{statistical classifiers} such as linear classifiers (e.g., Naive Bayes classifier or logistic regression), hidden Markov model (HMM) and Bayesian networks, or \emph{connectionist} classifiers such as neural networks.

A central result in `supervised learning theory' is the `\emph{no free lunch theorem}' which informs that there is no single learning method that will outperform all others regardless of the problem domain and the underlying distributions. For this reason, a variety of domain and application specific techniques have emerged to deal with diverse applications with varying degrees of success. The design of practical learning algorithms is therefore a mixture of art and science \cite{kulkarni2011elementary}.

\vspace{2mm}
\emph{Major issues in supervised learning:} The major issue with supervised learning is the need to generalize a function from the learned data so that the technique may be able to conjure up the correct output even for inputs it has not explicitly seen in the training data. This task of generalization cannot be solved exactly without some additional assumptions\footnote{These assumptions are subsumed in the phrase \emph{inductive bias}. See \cite{mitchell1997machine} for more details.} being made about the nature of the target function as it is possible for the yet unseen inputs to have arbitrary output values. Potential problems arise in supervised learning of creating a model that is underfitted (perhaps due to limited amounts of training data) or overfitted (in which a unnecessarily complex model is built to model the spurious and uncharacteristic noisy attributes of data).  Depending on the application, huge amounts of training data may be necessary for the supervised learning algorithm to work.

\subsection{\uppercase{Unsupervised learning}}
\label{unsupervised}

In supervised learning, it was assumed that a labeled set of training data consisting of some inputs and their corresponding outputs was provided. In contrast, in unsupervised learning, no such assumption is made. The objective of unsupervised learning is to identify the structure of the input data. To understand unsupervised learning, again imagine the machine that experiences a series of inputs: $x_1$, $x_2$, $x_3$, and so on. The goal of the machine in unsupervised learning is to build a model of \textbf{x} that can be useful for decision making, reasoning, prediction, communication,  etc.

The basic method in unsupervised learning is clustering (which can be thought of as the unsupervised counterpart of the supervised learning task of classification). This clustering is used to find the groups of inputs which have similarity in their characteristics.

\vspace{2mm}
\emph{Application of unsupervised learning to CRNs:} An application to which unsupervised learning is particularly suited to is the extraction of knowledge about primary signals on the basis of measurements \cite{bkassinysurvey}. A prominent unsupervised classification technique that has been applied to CRNs particularly for this problem is the Dirichlet process mixture model (DPMM). The DPMM is a Bayesian non-parametric model which makes very few assumptions about the distribution from which the data are drawn by using a Dirichlet process prior distribution \cite{teh2006hierarchical}. The benefit of Dirichlet process based learning is that training data is not needed anymore, thus allowing this approach to be used for identification of unknown signals in an unsupervised setting. Dirichlet process has been proposed in literature \cite{shetty2009identifying} for identifying and classifying spectrum usage by unidentified systems in CRNs.

\subsection{\uppercase{Reinforcement learning}}
\label{reinforcement}

Reinforcement learning (RL) is inspired from how learning takes place in animals. It is well known that an animal can be taught to respond in a desired way by rewarding and punishing it appropriately; conversely, it can be said that the animal \emph{learns} how it must act so as to maximize positive \emph{reinforcement} or reward. A crucial advantage of reinforcement learning over other learning approaches, and a main reason for its practical significance, is that it does not require any information about the environment except for the reinforcement signal.

To understand RL, we again take recourse to the example of the machine which experiences a series of inputs: $x_1$, $x_2$, $x_3$, and so on. In this new setting, the machine can also perform certain actions $a_1$, $a_2$, ... through which it can affect the state of the world and receive rewards (or punishments) $r_1$, $r_2$, and so on.\footnote{The reinforcement is a scalar value that can be negative to express a punishment or positive to indicate a reward.} The mapping from the actions to rewards is probabilistic in general. The objective of a reinforcement learner is to discover a \emph{policy} (i.e., a mapping from situations to actions) such that \emph{expected} long-term reward is maximized.

\section{\uppercase{Decision and Planning Techniques}}
\label{sec:decandplanning}

The cognitive cycle which epitomizes the essence of a cognitive radio is based on a cognitive radio's ability to: \emph{i)} \emph{observe} its operating environment, decide on how to \emph{ii)} best \emph{adapt} to the environment, and then as the cycle repeats, to \emph{iii)} \emph{reason} and \emph{iii)} \emph{learn} from past actions and observations \cite{gavrilovskalearning}. The term \emph{planning}, for the purpose of our discussion, refers to any computational process that produces (or improves) a decision \emph{policy} of how to interact with the environment given a model of the environment.  Planning is sometimes often referred to as a \emph{search} task, since we are essentially searching through the space of all possible plans \cite{mitchell1997machine} \cite{sutton1998reinforcement}.

In the remainder of this section, we will discuss two major decision planning frameworks that have been widely applied to CRNs. Specifically, we shall be studying Markov decision processes and game theory.

\subsection{\uppercase{Markov Decision Processes:}}
\label{sec:MDP}

Markov decision processes (MDPs) provide a mathematical framework for modeling sequential planning or decision making by an agent in real-life stochastic situations where the outcome does not follow deterministically from actions. In such cases, the output (also, called the reward) is specified by a probability distribution that depends on the action adopted in a particular state. MDPs approach this multi-stage decision making process sequential as an `optimal' control problem in which the aim is to select actions that maximize some measure of long-term reward.\footnote{Please see figure \ref{fig:markovrelation} and table \ref{tab:connections} to see how MDPs relate to other techniques and AI related fields.} MDPs differ from classical deterministic AI planning algorithms in that its action model is stochastic (i.e., the outcome does not follow deterministically from the action chosen).

More formally, an MDP is a discrete time stochastic optimal control process.  Every time step, the process is in some state $s$, and the decision maker has to choose some action $a$ from amongst the $A$ actions available in the current state.  After taking the action, the process will move randomly to some new state $s'$, with the decision maker obtaining a corresponding reward $R_a(s,s')$. We note here that the reward is used in a neutral sense: it can imply both a positive reward or a negative reinforcement (i.e., a penalty). The choice of action $a$ in state $s$ influences the probability that the process will move to some new state $s'$.  This probability (of going from state $s$ to $s'$ by taking action $a$) is given by the state transition function $P_a(s,s')$.\footnote{In some literature, the state transition function $P_a(s,s')$ is expressed through the alternative notation of $T(s,a,s')$.} The next state $s'$, therefore, depends stochastically on current state $s$ and the action $a$ taken therein by the decision maker.  In MDPs, an extra condition holds crucially:  given $s$ and $a$, the $P_a(s,s')$  is conditionally independent of all previous states and actions.  This condition is known as the \emph{Markov property} and this condition is critical for keeping MDP analysis tractable.

To put MDPs into perspective, we note here that they are a generalization of Markov chains. The difference is that MDPs incorporate actions and rewards in the model while Markov chains do not. Conversely, the special case of MDPs with only one action available for each state and with identical rewards (e.g., zero)  is in fact a Markov chain. This, and the relationship of various Markov models and games that we will develop later in this paper, can be seen graphically in figure \ref{fig:markovrelation}.

The roots of such problems can be traced to the work of Richard Bellman \cite{bellman1957} who showed that the computational burden of solving an MDP can be reduced quite dramatically via techniques that are now referred to as \emph{dynamic programming} (DP). We will discuss these techniques next.

\vspace{2mm}
\emph{Solving an MDP:} The core problem in MDPs is to determining an optimal `\emph{policy}' for the decision maker which is defined to be a function $\pi$ that maps a state $s$ to an action $\pi(s)$. Intuitively, the policy $\pi$ specifies what action must the agent perform when in various states so that the long-term rewards are maximized. It may be noted that once the MDP is specified with a policy, the action at various states is fixed, and the resulting MDP effectively behaves like a Markov chain.

We can now make the notion of long-term rewards more precise now. In a potentially infinite horizon environment, with continuous decision making which goes on forever, to reason about the various different possible policies, it is important that the reward function be non-finite. This is usually accomplished through \emph{discounting} through which the preference of immediate rewards over delayed rewards may be quantified. Discounting works by reducing future rewards by a factor of $\gamma$ chosen such that $0 \le \gamma < 1$ in every time step. The discount factor $\gamma$ is used as a parameter to describe the relative importance of future rewards. If $\gamma$ is chosen to be 0, the agent will become short sighted or \emph{`myopic'} and will consider current rewards only. As $\gamma$ approaches 1, the agent will become long-sighted and it will strive for long-term rewards. To ensure that action values do not diverge, the discount factor should not be equal to, or exceed, 1. Solving an MDP now entails determining the policy $\pi$ that maximizes the cumulative discounted reward function over a potentially infinite horizon: $\sum^{\infty}_{t=0} {\gamma^t R_{a_t} (s_t, s_{t+1})}$  where we choose $a_t = \pi(s_t)$, $\gamma$ is the discount factor, and the subscript $t$ refers to the time-step.

We can also define the \emph{value} of a state which follows naturally from the concept of rewards. Intuitively, the value of a state is a sum of discounted rewards that accrue from following the optimal policy onwards from that state. More precisely, $V(s)$ or the value of a state $s$ will contain the expected sum of discounted rewards to be earned (on average) by following the policy $\pi$ from state $s$. A \emph{value function} is a mapping from the states to their values or expected upcoming cumulative reward. For compactness, we refer to $R_{a_t} (s_t, s_{t+1})$ where $a_t = \pi(s_t)$, or the reward achieved in time $t+1$ by following the optimal policy $\pi$ at time $t$ simply as $r_{t+1}$. The value function mapping is shown below.

\begin{equation}
V(s_t) = E [ r_{t+1} + \gamma r_{t+2} +\gamma r_{t+3} + ... ]
\label{eq:extensive_value}
\end{equation}

It is worth emphasizing that the value abstraction is a key idea, and all efficient methods for solving sequential decision problems estimate value functions as an intermediate step \cite{valuefunction-sutton}. Apart from using the equation above (eq. \ref{eq:extensive_value}), another efficient, but remarkably simple, method can be used  for calculating the value function on the basis of \emph{bootstrapping}. We will see this method when we later will study eq. \ref{V(s)} when the \emph{Bellman equation} is introduced.

We emphasize again that it is due to the Markov property that the optimal policy $\pi$ is written as a function of only the current state $s$ and not of the past trajectory of the process through various states. We shall see later than analysis becomes intractable and convergence guarantees are lost when this condition is not met.

\vspace{2mm}
\subsubsection{Dynamic programming solutions to MDPs}

Assuming that we wish to calculate the policy that maximizes the expected discounted reward given that the state transition function $P$ and the reward function $R$ is known (this assumption is not always met, but we start with this simple case).

The naive approach to the problem of optimal sequential decision making would be to consider the set of all feasible policies, compute the return for each, and then to choose the policy providing the maximum return. This brute-force approach will not work except for the most trivial problems and will be hopelessly inadequate for processes involving even a moderate number of stages and actions. If we momentarily re-examine the situation practically, we will see that this price of excessive dimensionality arises from too much information. How much information is actually needed to carry out a multi-stage decision process?

The basic idea of the theory underlying dynamic programming is refreshingly simple. Optimal policy should be viewed as determining the decision required at each time in terms of the current state of the system.  Regardless of the initial state and decisions,  the remaining decisions must constitute an optimal policy $\pi$ for the continuation process treating the current state as starting input. This is known as the \emph{principle of optimality}. This strikingly simple insight allows computation of the optimal policy through backward induction starting at the terminal point. The concept of value function $V$ is related to this, and it captures the expected future utility at any node of the decision tree, if we assume that an \emph{optimal policy will be followed in the future}.

\vspace{2mm}
\emph{Value Iteration Algorithm:}
\label{sec:VI}
\vspace{2mm}

The standard method of calculating this optimal policy requires calculation of the value function and the policy function. These two functions are stored in two arrays indexed by state: \emph{ i)} value $V$ containing the real values of states, and \emph{ ii)} policy $\pi$ which contains the actions of states. At the end of the algorithm, $\pi$ will contain the optimal solution (i.e., actions to perform for each state) while $V(s)$ will contain the values of various states (capturing the expected discounted sum of the rewards to be earned by following the policy $\pi$ from that state).

The algorithm has the following two steps that are repeated for all the states until the values converge. These steps are defined recursively as follows. Note that the two equations above are intimately connected. In particular, the calculation of $V(s)$ utilizes current policy information from $\pi(s)$.

\begin{equation}
\pi(s) = \arg \max_a \left\{ \sum_{s'} P_a(s,s') \left( R_a(s,s') + \gamma V(s') \right) \right\}
\label{pi(s)}
\end{equation}

\begin{equation}
V(s) = \sum_{s'} P_{\pi(s)} (s,s') \left( R_{\pi(s)} (s,s') + \gamma V(s') \right)
\label{V(s)}
\end{equation}

Before discussing eq. \ref{V(s)} in more detail, it is contrasted with a method we have earlier derived for calculating $V(s)$ in eq. \ref{eq:extensive_value}. The method in eq. \ref{eq:extensive_value} was based on an explicit summation over expected future rewards. It turns out that eq. \ref{V(s)}, which also happens to be the \emph{Bellman equation} for this process, is considerably more simple and useful for practical purposes. The key insight here is to employ \emph{bootstrapping} to  estimate the values of states iteratively and recursively. This is done by relating the value of each state to the values of the states that follow it. The Bellman equation for calculating $V(s)$ can be alternatively expressed more simply as follows:

\begin{equation}
V(s_t) = E [ r_{t+1} + \gamma V(s_{t+1}) ]
\label{eq:bootstrapping_value}
\end{equation}

While both the definitions of calculating value functions (based on the extensive definition in eq. \ref{eq:extensive_value} and the bootstrapping definition in eq. \ref{eq:bootstrapping_value}) have the same exact solution, they tellingly have different approximate solutions. The bootstrapping eq. \ref{eq:bootstrapping_value}) is considerably more convenient in terms of time.

In value iteration, proposed by Bellman in 1957 \cite{bellman1957}, the policy function $\pi$ is not used directly. The value of $\pi(s)$ is instead calculated indirectly within $V(s)$ whenever it is needed. This technique is also known by the name backward induction. Substituting the calculation of $\pi(s)$ into the calculation of $V(s)$ gives us the following \emph{Bellman equation} for this problem. The value iteration update works by iteratively calculating the values of $V(s)$.

\begin{equation}
V(s) = \max_a \left\{ \sum_{s'} P_a(s,s') \left( R_a(s,s') + \gamma V(s') \right) \right\}
\label{eq:VI}
\end{equation}

Note that eq. \ref{eq:VI} is just an alternate representation of eq. \ref{V(s)} but it serves to emphasize a potential problem that can arise with value iteration when it comes to solving complex large-scale MDPs. For each action, we calculate a weighted average over possible outcomes to determine the expected reward from that action. We then choose the action with the maximum expected reward. Since the equation above is taking a maximum over \emph{all} possible actions, this calculation does not lend itself naturally to the usage of approximate methods. With the preclusion of approximation techniques, this method then becomes unwieldy for large-scale problems complex problems.

%We will see later in section \ref{sec:qlearning} how this problem is resolved by using methods like Q-learning which do not have %to calculate the expected rewards for all possible actions.  [CONFIRM]

%There are many variants of this basic algorithm. The exact order the two steps are followed through depends on the variant being %used. of the algorithm; one can also do them for all states at once or state by state, and more often to some states than %others. It has been shown that the algorithm will eventually arrive at the correct solution as long as no state is permanently %excluded from either of the steps.

\vspace{2mm}
\emph{Policy Iteration Algorithm:}
\label{sec:PI}
\vspace{2mm}

Policy iteration was devised based on the observation that it is possible to get an optimal policy even with inaccurate value function estimate or before this function converges. This is especially the case when one action is clearly better than all others; in such a case, it becomes clear what action needs to be taken even with imprecise estimates of the exact value magnitudes \cite{russell1995artificial}.

This insight can be exploited to devise a new strategy for calculating optimal policies called \emph{policy iteration algorithm} that directly explores the policy space. This algorithm begins from some initial policy $\pi_0$ and thereafter alternates between the following two steps:

\emph{1) Policy evaluation:} Given a policy $\pi_i$, calculate $V_i = V^{\pi_i}$ which calculates the value of each state if $\pi_i$ is to be executed.

\emph{2) Policy improvement:} Given $V_i$, calculate $\pi_{i+1}$ using one step look ahead based on $V_i$ (as in eq. \ref{pi(s)}).

The policy iteration algorithm terminates when the policy improvement step yields no change in the utilities.

The choice of which solution method is better depends on various factors. If there are many actions, or if there exists already a fair policy, it is better to use policy iteration. On the other hand, if there are few actions, and acyclic state transitions, then value iteration is a better option.

\vspace{2mm}
\emph{Partially observed MDPs:}

A MDP in which the environment is only partially observable is known as a partially observable MDP (POMDP). In the method discussed above for solving MDPs, it was assumed that the state $s$ is known when the action is to performed. This assumption does not hold for POMDPs. POMDPs are able to model uncertain aspects of the environment such as the stochastic effects of actions, incomplete information and noisy observations over the environment. Although POMDPs have been known for decades, their widespread uptake is impeded for two main reasons: i) it is difficult to satisfactorily model the environment dynamics (such as probabilities of action outcomes and the accuracy of data), and ii) it is difficult to solving the resulting model.

\vspace{2mm}
\subsubsection{Solutions for complex MDPs}

While the classical DP algorithms of value iteration and policy iteration work very well for simple to moderately complex MDPs, they break down for large-scale and complex MDPs as the requirement of computing, storing, and manipulating the so-called transition probability matrices becomes prohibitive. In complex MDPs, two crippling problems arise: \emph{i)} \emph{the curse of modeling}, and, \emph{ii)} \emph{the curse of dimensionality}. In the former problem, it becomes very difficult to compute the values of the transition probabilities while for the latter problem, storing or manipulating the elements of the so-called value function needed in DP becomes challenging due to the large dimensionality. Therefore, classical DP techniques are rather ineffective at solving large-scale complex MDPs \cite{gosavi2009reinforcement}.

\emph{Dealing with MDPs with unknown probabilities}: If the probabilities of MDP are unknown, then the problem becomes a reinforcement learning (RL) task. We have earlier seen RL in section \ref{reinforcement} where we noted that the task of RL is to determine for an agent what actions it should take in a stochastic environment. We will methods of dealing with this when we develop solutions for RL later in section \ref{sec:RL}.

\vspace{2mm}
\subsubsection{Previous work of applying MDPs in CRNs}

MDPs have been applied to study a wide range of planning and optimization problems in CRNs. It is noted here that MDPs in their native form require complete knowledge of the system (such as the state transition probabilities and the number of states, etc.) and they are not directly applicable when CRs are operating in unknown RF environments. However, various techniques exist (such as reinforcement learning) that can work in such scenarios where the environment is not completely known. In \cite{choi2011opportunistic}, Choi et al. proposed a partially observable Markov decision process (POMDP) based framework for channel access to opportunistically exploit frequency channels a primary network operates on. In another work, Zhao et al. had devised a POMDP framework to develop a cognitive MAC protocol \cite{zhao2007decentralized}. MDPs have also been applied extensively in communication networks. Interested readers are referred to a survey paper \cite{altman2002applications} which highlights the applications of MDPs to communication networks, and also includes a discussion on its use for routing.

\begin{figure*}[t]
\begin{center}
\includegraphics[width=.85\textwidth]{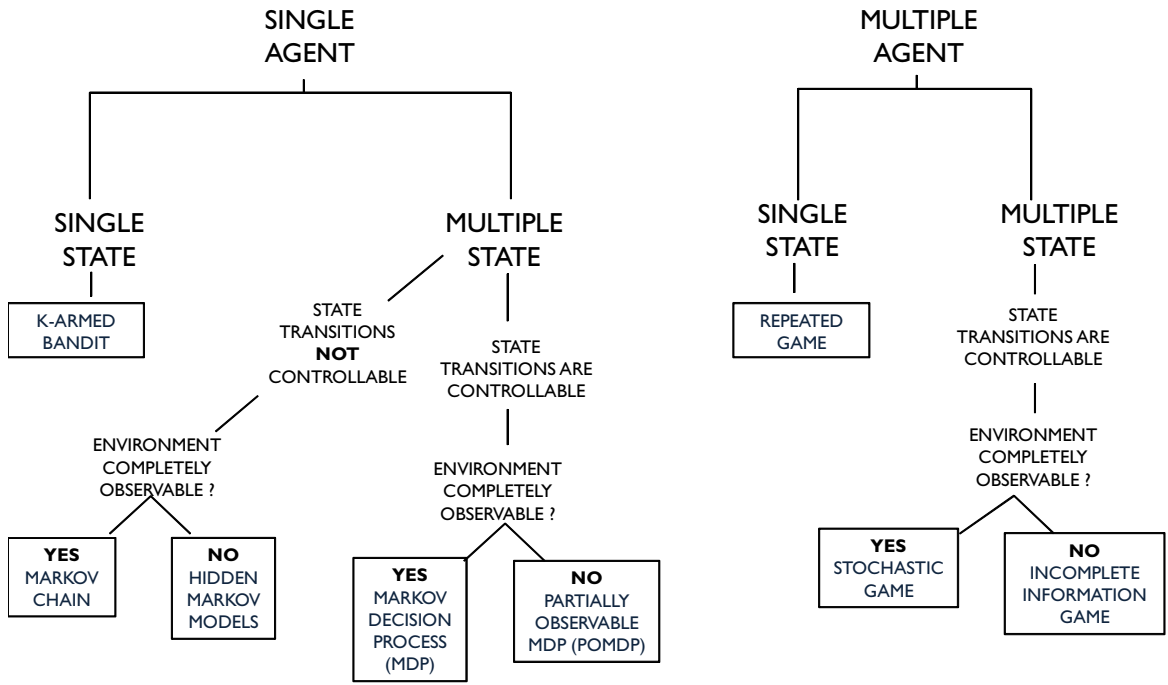}
\caption{Relationship between various Markov models, processes, and games.} \label{fig:markovrelation}
\end{center}
\end{figure*}

\subsection{\uppercase{Game Theory}}

Game theory is a mathematical decision framework composed of various models and tools through which we can study and analyze \emph{competitive} interaction between multiple self-interested \emph{rational} agents. Although, game theoretic models exist for both cooperative and non-cooperative settings, the ability to model competition mathematically distinguishes game theory from optimal control-theoretic frameworks such as the MDP \cite{haykin2005cognitive}. Game theory is also differentiated from optimization theory (which caters to a single decision maker scenario) in their ability to model \emph{multi-agent decision making} scenarios where the decisions of each agent affect each other.

Every \emph{game} involves a set of \emph{players}, \emph{actions} for each of the players representing how players interact, \emph{preferences} for each of the players defined over all the possible outcomes. The preferences, or \emph{payoffs}, are typically defined through a \emph{utility function}, or a \emph{payoff function}, which maps each possible outcome to a number representing that outcome's desirability. An outcome brings more reward, or is more desirable, if it has a higher utility \cite{mackenzie2006game}. In order to maximize its \emph{payoff}, each player acts according to its \emph{strategy}. More formally, a game can be mathematically represented by the 3-tuple $G = (N, S, U)$ where $N$ represents the set of players, $S$ the set of strategies, and $U$ the set of payoff functions.

The terms strategy and action should not be confused together: the strategy in fact specifies how the player should \emph{act} in each possible situation, and can be envisioned as a complete algorithm documenting how the player will play the game. The strategy of a player can be a single action (for a single-shot or a \emph{static} games) or a set of actions during the game (for a sequential or a \emph{dynamic} games) \cite{felegyhazi2006game}. A player's \emph{strategy set} defines what strategies are available for it to play: the strategy set may be finite (e.g., when a choice is made from a countable discrete set of values) or infinite (e.g., when some continuous value is chosen). A \emph{pure strategy} deterministically defines how a player will play a game, while a \emph{mixed strategy} defines a stochastic definition by assigning probability to each pure strategy. The \emph{strategy profile}, or the \emph{action profile}, documents the strategy of each player and it fully specifies all actions in a game. The outcome of the game depends, possibly stochastically, on the player's strategy profile and returns payoffs to various players.

Game theory is popularly used in CRNs since each CR in a CRN interacts with a dynamic environment composed of other \emph{rational} agents that sense, act, and learn while aiming to maximize personal utility. For games specific to CRNs, individual CRs typically represent the players, and the actions may include the choice of various system or design parameters such as, e.g., the modulation scheme, transmit power level, flow control parameter, etc. One of the main goal of game theory is to determine \emph{equilibria} points for a given game. These are sets of stable strategies in which individuals are unlikely to unilaterally change their behaviour. To gauge their efficiency, these equilibria points are often contrasted with some notion of socially \emph{optimal} point which produces the `best' outcome when interests of all the players is taken into accounts.

In recent years, game theory has provided deep insights into how to design decentralized algorithms for resource sharing in networks particularly through the theory known as \emph{mechanism design} sometimes known as reverse game theory. While traditional game theory focuses on analyzing how rational players would play a given game, in mechanism design, we are interested in engineering or design a game which rational players will play into a desired equilibrium point. Intuitively, mechanism design aims to set up the game such that players do what the designers want them to do but because the players themselves want to do it \cite{keshav2012mathematical}.

\vspace{2mm}
\subsubsection{Representation of games}

There are two common ways of representing non-cooperative games. The \emph{normal-form} representation of a game explicitly lists the payoff for each player of every conceivable outcome. This representation, also known as the \emph{standard-} or \emph{strategic-form}, is appropriate for static games of complete, and perfect information. For two player games, this can be depicted in a matrix form either as a pair of payoff matrices (one each for the \emph{row player} and \emph{column player}) or as a single payoff matrix (with an entry containing payoffs for both players). On the other hand, an \emph{extensive-form} game is a representation that allows, unlike the normal-form games, explicit representation of temporal aspects of dynamic games such as the sequencing of players' possible moves and their choices at every decision point along with payoffs for all possible game outcomes. It also allows representation of the (possibly imperfect) information each player has about the other player's moves when making a decision, and of incomplete information (about the nature of the game) in the form of chance events encoded as moves by the player `nature'. More details about representation of the games can be seen at \cite{mackenzie2006game}.

\vspace{2mm}
\subsubsection{Solution Concepts}

In game theory, a \emph{solution concept} formalizes the concept of `solving' a game by predicting how rational players would play a specified game. These predictions, called \emph{solutions}, describe what strategies would be chosen by players and, therefore, it also describes the predicted result of the game. The most commonly used solution concepts are equilibrium concepts and the optimality concepts.

%There are two aspects that relate to solving a game: \emph{i) descriptive} aspect that tries to describe or predict how the players will actually play the game, and \emph{ii): prescriptive} aspect that tries to prescribe the way players should play given the game specifications.

We shall now discuss three concepts of equilibrium that are relevant to our subject.

The \emph{Nash equilibrium} (NE) is a solution concept of a non-cooperative game involving two or more players. A NE is a stable equilibrium point of a game representing the situation where no player can benefit by changing its strategy unilaterally (i.e., by the player changing its strategy while other players keep their unchanged). In other words, a NE implies that each player's strategy is the \emph{best response} against those of the others. It is noted that it is possible for games to have multiple NE. While NE is a very useful concept, analysis based solely on NE has many drawbacks as pointed out in \cite{haykin2005cognitive} \cite{halpern2008beyond}. Also, the significant complexity of computing NEs has prompted development of alternative solution concepts.

The \emph{Correlated equilibrium} is an intuitive solution concept that generalizes the Nash equilibrium and is much easier to compute.\footnote{Roger Myerson has pithily remarked that: ``If there is intelligent life on other planets, in a majority of them, they would have discovered correlated equilibrium before Nash equilibrium.''} The idea is that each player chooses its action after observing a common public signal. The player's strategy assigns an action to every possible observation. If no player has any incentive to deviate from the devised strategy, assuming that others don't deviate, the game is in correlated equilibrium.

The \emph{Wardrop equilibrium} is a common solution concept useful for modeling selfish routing in transportation and telecommunication networks with congestion. It is assumed that in the study of transportation and telecommunication networks that the players (travelers or packets, respectively) choose the shortest perceived routes given the current traffic conditions. For a network in Wardrop equilibrium, all the flow paths in use for a source-destination pair have an equal delay. No other unutilized path has a lower delay in the Wardrop equilibrium.\footnote{If this property was not met, the system would not be in equilibrium intuitively, for it would have been possible for a flow to reduce its latency by switching to an unutilized path.} A wireless routing analogue of this was explored in \cite{raghunathan2009wardrop} where a flow-avoiding routing protocol was proposed.

While \emph{optimality} has a well-defined unambiguous meaning in optimal control problems (one-player games), optimality, in settings of multi-player decision making, is a difficult concept to define precisely. Equilibrium points are not necessarily optimal since equilibria points may not be `socially optimum' (e.g., as in the classical Prisoner's dilemma game \cite{nisan2007algorithmic}). A common notion of optimality in game-theory is that of \emph{Pareto-optimality}. A strategy profile is stated to be a \emph{Pareto-optimal} solution if no other joint decision of the players can improve the performance of at least one of them without degrading the performance of another. It must be noted that achieving Pareto optimality does not imply equality nor fairness. Another optimality concept is the \emph{Minimax} solution concept useful for non-zero-sum games in which it is aimed to minimize the maximum loss a player will face in the worst-case scenario \cite{basar1995dynamic}.

Game theory predicts the agents' equilibrium behavior typically without specifying by itself how to reach such a state. Algorithms for computing equilibria and determining the dynamics of games towards it is a subject studied in the fledgling discipline of \emph{algorithmic game theory} which is at the intersection of game theory and algorithms \cite{nisan2007algorithmic}. It has been shown that equilibrium points do not have necessarily have to socially optimal. An interesting question then is to quantify how inefficient the equilibria points (which are reached through self interested behavior) are with reference to the idealized `optimal' situation (where the agents collaborate selflessly in a bid to minimize total cost). Since there can be multiple NE with varying overall payoffs, the comparison of the worst NE with the ideal is known as the \emph{`price of anarchy'} while the comparison of the best NE with the ideal is known as the \emph{`price of stability'} \cite{nisan2007algorithmic}.

We have covered only the most basic solution concepts that are relevant to our subject. For a discussion on advanced solution concepts such as rationalizability, $\epsilon$-Nash equilibrium, trembling-hand perfect equilibrium, we refer the interested reader to standard game theory texts \cite{leyton2008essentials}.

\vspace{2mm}
\subsubsection{Categories of games}

There are various ways to categorize games, we will discuss games through the following six contrasting categories:

\begin{enumerate}

\vspace{2mm}
\item \emph{Cooperative vs. non-cooperative:} in all game theoretic models, a basic primitive is the concept of a \emph{player}. A player may be either be interpreted as an individual or alternatively as a group of individuals. After defining the set of players in a game, we may distinguish between two kinds of models: \emph{i)} in which we are dealing with the possible actions of individual players; \emph{ii)} in which we are dealing with possible joint actions of groups of players. Models of the former kind (individual-based) are sometimes known as `noncooperative', while those of the latter kind are correspondingly known as `cooperative'. The difference can be summarized in that in a cooperative game, players can make binding commitments, while in noncooperative game, they cannot. A game in which the players are groups of individuals that can make binding commitments is also known as a coalition game \cite{saad2012cooperative}.

\vspace{2mm}
\item \emph{Complete vs. incomplete information:} A game with complete information is a game in which each player knows the exact game being played. The game is represented by 3-tuple $G = (N, S, U)$ with $N$ representing the set of players, $S$ the set of strategies, and $U$ the set of payoff functions. This complete information is not known in games of incomplete information. We typically employ the model of a \emph{Bayesian game} to model situations in which some of the parties are not certain of the characteristics of some of the other parties. Games with incomplete information should not be confused with games with imperfect information (in which the history of the game is not available to all players). In a Bayesian game, at least one player is unsure of the type (and therefore the payoff function) of another player. In games of imperfect information, on the other hand, while the actual moves of agents are not common knowledge, but the game itself is.

\vspace{2mm}
\item \emph{Sequential vs. simultaneous:}  In a \emph{sequential game}, one player chooses his action before the others choose theirs---the latter player can utilize knowledge about the previous move to decide on its action. In \emph{simultaneous games}, on the other hand, players choose their moves without being aware of other player's moves. A game in which players have sequential interaction is also known as a \emph{dynamic game}.

\vspace{2mm}
\item \emph{Static vs. dynamic:} In \emph{static games}, alternatively known as \emph{single-stage games} or \emph{one-shot games}, it is assumed that there exists only a single time step implying that the players only have one move as a strategy. However, in a \emph{dynamic game}, players interact with each other sequentially. \emph{Repeated games}, also known as supergames, are a subclass of dynamic games in which a similar stage game is played numerous times. Players in a repeated game, unlike those in simultaneous games, have the benefit of historic information which they can utilize to adapt their strategy. Depending on the number of stages, we can classify dynamic games into \emph{finite-horizon games} and \emph{infinite-horizon game}---the strategies for such games can hugely vary. If players in a finite-horizon game are not aware of the duration of the game (which is clearly a common situation in practical interactions particularly in a networking setting), then infinite-horizon games with \emph{discounting} can be used an appropriate model. In order to cater for the potentially abrupt end to the game, discounting entails decreasing the value of future stage payoffs so that payoffs in nearer-by time are preferred. The study of dynamic game is taken in a subfield of game theory known as \emph{dynamic game theory} which can be envisioned as child discipline of game-theory and optimal control theory \cite{basar1995dynamic}.

\vspace{2mm}
\item \emph{Perfect vs. imperfect information:} We refer to a game as a \emph{perfect-information game} if the players have perfect knowledge of all previous moves in the game at any moment they have to make a new move. Since players in simultaneous games (which includes practical games like poker and bridge) do not know the actions of other players, simultaneous games are \emph{imperfect-information games}. Only sequential games, therefore, can be games of perfect information, with an an example sequential perfect-information game being chess.

\vspace{2mm}
\item \emph{Symmetric vs. asymmetric:} If the game is symmetric, the identities of the players may be changed without changing the payoff to the strategies. In other words, even if the role of the two players in a two-player symmetric game is reversed, the same payoffs would be observed. This condition does not hold for asymmetric games.

\vspace{2mm}
\item \emph{Zero-sum vs. non-zero-sum:} In a \emph{zero-sum game}, the sum of payoffs of all the players must be zero---in other words, a player cannot get better off without affecting some other player's utility. A game which is not zero-sum is called \emph{nonzero-sum game} or \emph{variable-sum game}.

\end{enumerate}

Uncertainty can come into games in three distinct ways: \emph{i)} a player may use chance to determine which strategy to use (such a strategy is known as mixed strategy), \emph{ii)} the game itself can include random events, and \emph{iii)} you may not be exactly sure what game you're playing---i.e., you may not know what strategies other players are capable of, or their payoffs precisely. The latter two points refer to the \emph{incomplete information} nature of the game. In addition, the game may have \emph{imperfect information} where the players do not know previous history or have \emph{asymmetric information}. We note here that simultaneous games are always imperfect information games since players choose their moves without being aware of other player's moves.

\vspace{2mm}
\emph{Stochastic games}, introduced by Lloyd Shapley in 1950s, are games in which (potentially multiple) agents take decisions in a sequence of stages (i.e., in a dynamic game) and each player receives a payoff that depends probabilistically on the current state and the chosen actions \cite{haykin2005cognitive}. Intuitively speaking, the agents in a stochastic game repeatedly play games from a collection of games---the particular game played at any given iteration depends probabilistically on the previous game played and on the actions taken by all agents therein \cite{leyton2008essentials}. Stochastic games have been applied in wireless networks in areas such as flow control, routing, and scheduling \cite{hossain2009dynamic}.

Stochastic games generalize the concepts of MDPs, Markov chains and repeated games---MDPs can be viewed as the special case of a single-agent stochastic game, Markov chains as single agent stochastic game where each player has a single action in each stage, while repeated games can be viewed as a single state (or, single stage) stochastic game\cite{neyman2003stochastic}. We have seen previously that MDP are appropriate models for reinforcement learning techniques that address the problem of a single agent learning through experience and interaction with an environment (assumed stationary). Stochastic games extend the concept of MDPs for multi-agent environments. In multi-agent environments, the other agents are also learning and adapting and thus the environment can no longer be assumed stationary. Stochastic games, also called competitive MDPs, allow us to model uncertainty in the players' operating environment by allowing probabilistic state transitions in a dynamic game.

\vspace{2mm}
\emph{Auctions:} With a plethora of heterogeneous technologies, the wireless communication system has become quite complex. The dynamism of the overall wireless ecosystem has led researchers to explore using models from other similarly complex domains so that complementary mechanisms may be exploited. Indeed, there has been a lot of work in applying various economics-based approaches to wireless networking \cite{maharjan2011economic}. CRNs, in their distributed nature, complexity and heterogeneity, have become analogous to real-world markets \cite{zhangauction} and are amenable to incorporation of market mechanisms and incentives. \emph{ Auction theory} is an interdisciplinary field that has shown itself to be particularly useful for CRN applications. Traditional static methods of managing spectrum are grossly inadequate for modern CRNs, and the market mechanism of auctions seems to be a promising approach for distributed allocation of network resources. A detailed survey of various auction approaches for resource allocation in wireless networks is provided in \cite{zhangauction}.

Incidentally, there are clear connections between MDPs and game theoretic models, in particular stochastic games. The relationship between Markov Chains, MDPs, POMDPs, and HMM and Markov (or stochastic) games can be seen in figure \ref{fig:markovrelation}. MDPs are observable stochastic environments in which a single agent takes a decision by choosing an action given knowledge of the current state. Markov games, or stochastic games, generalize the MDP model to allow a pair of agents to control state transitions (either jointly or in alternation). Note that a one-state stochastic game is equivalent to an (infinitely) repeated game, while the special case of an one-agent stochastic game is equivalent to an MDP. A POMDP models partially observable stochastic environments in which a \emph{single agent} takes a decision while being provided with partial knowledge of the current state. In incomplete information games, on the other hand, \emph{multiple agents} control the transitions in the environment while having incomplete knowledge of the environment's state.

\begin{table*}
\caption{Summary of the various decision and planning techniques discussed in section \ref{sec:decandplanning}}
\centering
\begin{tabular}{l p{6cm} p{7cm}}
\toprule
 \textbf{\emph{Decision techniques}} &  \textbf{\emph{Application to CRNs}} &  \textbf{\emph{Application to Routing}}\\
%\vspace{2mm}
\midrule
   %\\
   \textbf{\emph{Markov Decision Processes}} & \emph{Opportunistic spectrum access}: \cite{choi2011opportunistic}; & \emph{Routing in ad-hoc CRNs} \cite{di2010learning};\\
   &\emph{Medium Access Control (MAC)}: \cite{zhao2007decentralized};& \emph{Routing in communication networks}: see references in \cite{altman2002applications}.\\
   &\emph{Cooperative spectrum selection}: \cite{di2011learning};&   \\
   %\\
   \textbf{\emph{Game Theory}} &
  \emph{Resource allocation}: see references in \cite{maharjan2011economic} \cite{zhangauction};   & \emph{Routing games} \cite{roughgarden2007routing} \cite{pavlidou2008game} \cite{han2011repeated};\\
 & \emph{Spectrum Sharing}: \cite{han2012game} \cite{van2009spectrum}; &   \emph{Mitigating selfish routing} \cite{felegyhazi2006nash}\cite{eidenbenz2005commit} \cite{wang2004truthful};\\
  & \emph{Medium Access Control (MAC)}: \cite{akkarajitsakul2011game};  &   \emph{Modeling routing}: see references in \cite{mackenzie2006game}. \\
%\vspace{2mm}
  & \emph{Security}: see references in \cite{liu2010cognitive}.\\
 \bottomrule
\end{tabular}
\label{tab:decandplanning}
\end{table*}

\vspace{2mm}
\subsubsection{Game theory for Wireless Networks}

There has been a lot of work in applying game-theoretic ideas to the design and analysis of wireless networks \cite{mackenzie2006game} \cite{han2012game} \cite{srivastava2005using} \cite{naserian2009game} and cognitive radio networks \cite{wang2010game}. A comprehensive survey of game-theoretic approaches developed for different multiple access schemes in wireless networks is provided in \cite{akkarajitsakul2011game}.

In \cite{felegyhazi2006game}, Felegyhazi present a tutorial on the application of game-theory in wireless networks. To clarify the concepts, four games are constructed for wireless networks that are analogous to classical games in game-theory literature. In particular, they proposed two games, the `Forwarder's dilemma' and the `Joint Packet Forwarding', that relate to network-layer issues of packet forwarding \cite{felegyhazi2006game}. The `Forwarder's dilemma' is analogous to the classical game-theoretic problem of `\emph{Prisoner's dilemma}' \cite{nisan2007algorithmic} in which \emph{iterated strict dominance} solution exists. It is shown that the Forwarder's dilemma problem is a symmetric nonzero-sum game, because the players can increase their payoffs by mutually cooperating. In the second problem of `Joint Packet Forwarding', no iterated strict dominant solution exists and therefore analysis in terms of \emph{Nash equilibrium} (NE) is shown---since this game has two NE, the example is exploited to explain the concept of Pareto optimality. The Joint Packet Forwarding problem is also nonzero-sum but it is no longer symmetric but is asymmetric.

Challenges and experiences in applying game-theoretic ideas to system design are related in \cite{mahajan2004experiences}.  Various approaches for incentivizing cooperative forwarding behavior were analyzed including bartering primitive, virtual currency primitive, and setting up a equilibrium point at a  desired forwarding rate through appropriate game mechanism design.

\vspace{2mm}
\emph{Application of game-theory in CRNs:} There is a lot of literature on the applications of game theory to CRNs. Interested readers are referred to the following two survey papers and a book and the references therein for more details. Van der Schaar presented a survey of spectrum-access games that are relevant to DSA CRN in \cite{van2009spectrum}, while a more general survey paper on game-theoretic ideas to CRNs was published by Wang et al. \cite{wang2010game}. A comprehensive game-theoretic treatment of cognitive radio networking and security is presented in the book authored by Liu et al. \cite{liu2010cognitive}.

\vspace{2mm}
\subsubsection{Game theory for Routing}
\label{sec:gametheoryrouting}

The framework of game-theory has presented itself as a viable choice for modeling the problem of routing in a network with some applications being identifying and mitigating selfish routing behavior, convergence of routing techniques with changing network conditions, and the effects of different kinds of node behavior on routing \cite{mackenzie2006game}. Some example works can be seen in the references of \cite{mackenzie2006game}.

An important aspect of tackling routing problems through game theory is precisely how the game is modeled (i.e., how are the players defined, what are the utilities, etc.). This is true of mathematical modeling in general where it is understood that models are mere abstractions of the reality being modeled and the purpose of models is to be useful rather than to be accurate.\footnote{The statistician George Box famously remarked that ``all models are wrong, some are useful''.} Various implications of how to model a problem of routing in network is discussed in \cite{mackenzie2006game}. To summarize the discussion in \cite{mackenzie2006game}, assume a simple \emph{source routing} setup (where the end-to-end path is specified by the source node), chosen for ease of exposition, the \emph{players} in the game can be viewed as the source nodes in the network, although, it can be more convenient to view a player as a source/destination pair (since such a formulation can allow for the existence of multiple flows from a single source.) The \emph{action} set available to each player is possibly the set of all possible paths from the source to the destination. Depending on how the game is formulated, a node may choose a single path from all the possible paths or even choose multiple paths and also how much of their flow to send on each route. \emph{Preferences} in a routing game can take several forms just like many routing metrics exist for routing protocols to determine a route's quality. A simple way to formulate preferences can be to base it on end-to-end delay for a packet to traverse the chosen route with a short delay being preferable to longer delay. While such a simple example can be solved through optimization techniques (especially, if we consider a single source and destination pair or if the available routes are completely disjoint), the benefit of using game theory kicks in when we consider the interaction between multiple flows using common paths through the network.

An interesting aspect of game-theoretic models of network problems is that it can explain certain nonintuitive behavior. For example, it has been shown that in certain cases, adding more resources (e.g., adding extra links) to a network in equilibrium can actually lead to a new equilibrium in which all the users are worse off. This phenomena, known as \emph{Braess' paradox} \cite{braess2005paradox}, shows that how the dynamic interaction between players and resources can lead to counterintuitive results and why using a mathematical theory like game theory can be a useful tool. Network routing problems also arises in domains other than telecommunication networks (e.g., transportation networks) have been studied for a long time with a common solution concept known as \emph{Wardrop equilibria} which has been discussed earlier.

In algorithmic game theory, \emph{selfish-routing} in networks is a well-studied problem both in a general network setting (e.g., of transportation networks) \cite{roughgarden2007routing} and also for Internet-like networks \cite{qiu2006selfish}. In general, centralized calculation of optimal routes are infeasible for a majority of network routing problems, leading to interest in distributed algorithms. Distributed algorithms can be viewed as `selfish routing' since each agent intends to optimize for itself. Researchers have vigorously pursued questions that aim to quantify the performance degradation due to lack of coordination between the various `players' of this \emph{routing game}.  In this regard, concepts of price of anarchy and price of stability, discussed earlier, have been proposed. It has been shown that while the price of anarchy is unbounded for the case of selfish routing in networks with general latency functions \cite{roughgarden2007routing}, results are much more encouraging for networks with linear latency functions \cite{roughgarden2007routing} and for actual Internet-like networks \cite{qiu2006selfish}. Selfish routing in networks and their equilibria was first formally defined by Wardrop in 1952, and it has been an popular topic for researchers since.

\vspace{2mm}
\subsubsection{Routing Games}
\label{sec:routinggames}

A characteristic of a typical routing game is that each player is interested in finding a minimum cost path from the origin to the destination in a \emph{congested} network, where the delay of an edge on some path depends on its congestion which in turn depends on the total of players using that edge in their path. Such a dependence on congestion is seen in a class of games known as \emph{congestion games}, first proposed by Rosenthal in 1973. In a congestion game, the payoff of each player depends  not only on the resource it chooses, but also on the number of players choosing the same resource. Congestion games are a special case of \emph{potential games}. Fortunately, the equilibria points are guaranteed to be approximately optimal under best response dynamics \cite{nisan2007algorithmic} for potential games in general.

\emph{Repeated games} and \emph{potential games} have been shown to be especially relevant to the routing problem. In previous work, repeated games have been used to address the problem of selfish routing with punishment for unsocial behavior \cite{felegyhazi2006nash}\cite{eidenbenz2005commit}\cite{wang2004truthful}. The usage of potential games for routing has been well-explored \cite{roughgarden2007routing}. Potential games encompass many of the well-studied network routing and congestion games. Potential games have many desirable properties including i) pure equilibria always exists, ii) the best response dynamics is guaranteed to converge, and iii) the price of stability (or, the ratio of the best NE to the optimal solution) can be bounded using a technique named the potential function method. Potential games are especially attractive from the point of view of analysis, since the incentives of all the players are mapped onto a single function, called the potential function, whose local optima correspond to the set of pure NE. There has been a lot of work in modeling wireless networking problems as potential games (see the references in \cite{hossain2009dynamic} for more details) with most applications being in the domain of power control, waveform adaptation, and routing and congestion games.

Broadly speaking, there are two popular models of routing games: \emph{nonatomic selfish routing} in which there are very large number of players each controlling a negligible fraction of overall traffic, and \emph{atomic selfish routing} in which each player controls a non-negligible amount of traffic. Nonatomic selfish routing was first studied for transportation networks by Wardrop, and equilibrium in such games is known as Wardrop equilibria. It has been shown that for nonatomic selfish routing, the price of anarchy is the same as the price of stability .  Nonatomic selfish routing has been applied to routing in communication networks where it is relevant to the `source routing' paradigm in which the source node specifies a complete route for its traffic  and in a distributed setting \cite{roughgarden2007routing}. The paradigm of distributed shortest-path routing, that is typically used on Internet-like networks, cannot be addressed by selfish routing unless the `length' used to define the shortest paths coincide with the edge cost functions \cite{roughgarden2007routing}.  Atomic selfish routing games were first considered by Rosenthal in 1973 who also introduced the concept of congestion games and potential games.  The price of anarchy is also well understood for atomic selfish routing game \cite{roughgarden2007routing}.

Interested readers are referred to a detailed survey of game-theoretic methodologies for routing models at \cite{pavlidou2008game}, details about \emph{routing games} and the analysis of the efficiency of its equilibria points at \cite{roughgarden2007routing}, and a survey of application of various networking games in telecommunications in \cite{altman2006survey}.

\section{\uppercase{Learning Techniques}}
\label{sec:learning}

Learning is especially crucial when dealing with unknowns or unplanned scenarios and is especially relevant to CRNs \cite{haykin2005cognitive}. Learning, for the purpose of our discussion, will focus on computational processes employed by CRs that can improve their behavior through diligent study of their own interactions with the environment. Learning can also be envisioned in the perspective of search. In this context, we can envision learning as searching through a space of possible hypotheses to determine which hypothesis best fits the available training examples and prior knowledge and constraints \cite{mitchell1997machine}.

In the remainder of this section, we will discuss hidden Markov models, reinforcement learning, learning with game theory, online learning algorithms, neural networks, evolutionary algorithms, support vector machines, and finally methods of Bayesian inference.

\subsection{\uppercase{Hidden Markov Models}}

Hidden Markov Model (HMM) are stochastic models of great utility, especially in domains where we wish to analyze temporal or dynamic processes such as speech recognition, PU arrival pattern in CRNs, etc. HMMs are highly relevant to CRNs since many environmental parameters in CRNs are not directly observable.

An HMM-based approach can analytically model Markovian stochastic processes whose actual states are hidden, but which emit observations from states per some probability distribution. It is for this reason that an HMM is defined to be a doubly stochastic process: first, the underlying stochastic process that is not observable, and second, the set of stochastic processes, dependent on the embedded underlying stochastic process, that produce the sequence of observed symbols \cite{rabiner1986introduction}.

Intuitively, HMMs can be visualized as a Markov chain observed in noise \cite{cappe2005inference}. In a simple Markov model like a Markov chain, the state is directly visible to the observer, and the model is completely specified by describing the parameters defined through state transition probabilities. In an HMM, on the other hand, a more elaborate model is needed. The relationship of HMM with other Markov models is depicted in figure \ref{fig:markovrelation}.

To represent an HMM, we use the notation $\lambda = (A,B,\pi)$ to represent an HMM where $A$, $B$ and $\pi$ are three probability distributions. $A$ is the \emph{state} transition probability, $B$ is the \emph{observation symbol} probability distribution from various states \cite{rabiner1986introduction}, while $\pi$ is the initial state distribution. Specifying an HMM completely requires, in addition to $A$, $B$ and $\pi$, information about the number of states $N$ and the number of discrete output symbols $M$.

%In terms of relative importance, $\pi$, or the initial conditions, is generally the least important while $B$, or the %observation symbol probability distribution from various states is the most important \cite{rabiner1986introduction}; $A$, or %the state transition probability, can also be important in certain problems \cite{rabiner1986introduction}.

\vspace{2mm}
\subsubsection{Key problems in HMMs}

Having defined the notation for HMMs above, we can talk about the three key problems that must be solved for the HMM to be useful in real world applications \cite{he2010survey} \cite{rabiner1986introduction}. The listing of these three keys problems below assumes an observation sequence $O = O_1, O_2, O_3, ... O_T$.

\begin{itemize}

\item \emph{Evaluation Problem}:  Given the parameters of the model $\lambda$, this problem deals with how to compute the probability of a particular observation sequence $Pr(O|\lambda)$. The forward algorithm, backward algorithm, and the forward-backward algorithm solve this problem.\footnote{While the forward-backward algorithm solve the evaluation problem (i.e., it can estimate the most likely state for any point in time), it cannot solve the decoding problem (of finding the most likely \emph{sequence} of states) for which the Viterbi algorithm is used.}

\item \emph{Decoding Problem}: Given the observation sequence $O$ and the parameters of the model $\lambda$, this problem deals with decoding or inferring about the sequence of hidden states $I = i_1, i_2, i_3, ... i_T$ that most likely produced the observation sequence. This task aims at decoding, or uncovering, the hidden part of the HMM and is essentially an estimation problem. The \emph{Viterbi algorithm} solves this problem by providing the most likely sequence and its probability.

\item \emph{Learning Problem}: Given an observation sequence $O$, this problem deals with learning the most appropriate model $\lambda= (A, B, \pi)$ that `best' explains the observed sequence. In other words, we have to learn the most likely set of state transition $A$ and observation symbol probabilities $B$ from the training data. For many applications, this is the most important task since it allows us to optimally adapt model parameters to the training data. The \emph{Baum-Welch} expectation-maximization algorithm solves this problem.

The learning problem in HMMs is intuitively related to evaluation problem in the following way. The evaluation problem computed $Pr(O|\lambda)$ which represented the probability of a particular observation sequence given a model. $Pr(O|\lambda)$ is also the likelihood function for $\lambda$ given the observations $O$. The learning problem is determining the HMM parameters $\lambda$ that maximize the likelihood function. The Baum-Welch algorithm is an iterative algorithm which solves the learning problem by expectation-maximization to produce maximum likelihood, or maximum a posteriori, estimates of HMM parameters given only observation sequence as training data.

\end{itemize}

We have already noted that HMM is a strong generic temporal model for dynamic signals and systems. To hone onto the important problem of inference in such temporal models, we note that there are four basic inference tasks that may be performed with HMMs \cite{russell1995artificial}. (We use the notation $I_t$ and $O_t$ to indicate respectively the hidden state and the observation during time step $t$. It is assumed that observations $O_0, O_1, ... , O_{t-1}$ have been observed till date.)

\emph{a) Filtering or Monitoring:} This is the task of computing the posterior distribution over the \emph{current state}, given all evidence to date. Mathematically, this is calculating $P(I_{t-~1}|O_0, ..., O_{t-1})$

\emph{b) Prediction:} This is the task of computing the posterior distribution over the \emph{future state}, given all evidence to date. Mathematically, this is calculating $P(I_t|O_0, ..., O_{t-1})$

\emph{c) Smoothing or Hindsight:} This is the task of computing the posterior distribution over \emph{past states}, given all evidence up to present.  Mathematically, this is calculating $P(I_k|O_0, ..., O_{t-1})$ for  $0 \le k \le t-1$

\emph{d) Most Likely Explanation:} This is the task mentioned earlier as the \emph{decoding} task.  The aim is to find the most likely sequence of states  that generated the observed sequence. Mathematically, this is
$\operatorname*{{arg\,max}_I}_{1:t} Pr(I_{1:t}|O_{1:t})$
%$\argmax_i_{1:t} Pr(i_{1:t}|O_{1:t})$

\vspace{2mm}
\subsubsection{Applying HMMs in CRNs}

All the inference tasks listed above are potentially very useful for CRNs. HMMs have been extensively used in CRNs for a wide range of problems. They can be used for spectrum prediction, PU detection, signal classification, etc. \cite{he2010survey}. A potential drawback when using HMMs is that a training sequence is needed, with the training process being potentially computationally complex. Other AI techniques such as GA are used to improve the model training efficiency \cite{rondeau2004online}. We will further discuss the usage of HMMs in section \ref{mod-and-pred-tasks} where we will outline how HMM has been, or can be, used for solving certain modeling, planning and prediction tasks that relate to cognitive routing in CRNs.

\subsection{\uppercase{Reinforcement learning}}
\label{sec:RL}

In reinforcement learning (RL), an agent aims to determine a sequence of actions or \emph{policy} which maps the state of an \emph{unknown stochastic} environment to an optimal action plan. We note here that MDPs, on the other hand, address this planning problem for \emph{known stochastic} environments. Since RL agents work in a stochastic environment, they have to balance two potentially conflicting considerations: on the one hand, it needs to \emph{explore} the feasible actions and their consequences (to ensure that it does not get stuck in a rut) while on the other hand, it needs to \emph{exploit} the knowledge, attained through past experience, of favorable actions which received the most positive reinforcement.

RL is distinct from supervised learning in that instead of being presented with training examples of how to select the correct output for an input, the system has to learn indirectly from reinforcements (called reward for positive reinforcement and punishment for negative reinforcement) on actions taken. Since reinforcement learning can be used without training data and because it aims to maximize the long-term online performance, it is particularly suitable for CRNs. Reinforcement learning is also distinct from supervised and unsupervised learning in that it focuses on online performance (\emph{learning through taking actions}) rather than on \emph{planning} and offline performance. Since it programs agents by reward and punishment without needing to specify how the task is to be achieved, and due to its broad applicability, the RL framework is of profound interest to many diverse fields.

We note here that reinforcement learning is also known by alternate monikers such as neuro-dynamic programming (NDP) \cite{bertsekas1995neuro} and adaptive (or approximate) dynamic programming (ADP) \cite{bertsekas2011approximate} \cite{gosavi2009reinforcement}.

\vspace{2mm}
\subsubsection{Relationship with MDPs}

An interesting way to conceptualize RL is to think of it as a simulation-based technique for solving large-scale and complex MDPs. We refer to section \ref{sec:MDP} for an earlier discussion on the relationship between MDPs and RL. We also discussed in section \ref{sec:MDP} that classical DP techniques are ineffective at solving large-scale complex MDPs \cite{gosavi2009reinforcement} \cite{szepesvari2010algorithms}. Practical RL algorithms that can deal with large-scale complex MDPs (having large state and action spaces) essentially bank upon two key ideas: firstly, to use samples to compactly represent the dynamics of the control problem, and secondly, to use powerful function approximation methods, including bootstrap methods that build estimates on other estimates, to compactly represent value functions \cite{valuefunction-sutton} \cite{szepesvari2010algorithms}. It has been stated that understanding the interplay between dynamic programming, samples and function approximation is at the heart of design, analysis and application of modern RL algorithms \cite{szepesvari2010algorithms}.

Crucially, RL can solve MDPs without explicit specification of the transition probabilities. These values are needed by classical dynamic programming solutions of value and policy iteration. In RL, instead of explicit specification of the transition probabilities, the transition probabilities can be envisioned to be accessed through a simulator that typically is restarted from a uniformly random initial state many times \cite{busoniu2011approximate}. In addition, RL can work with very large number of states when used along with function approximation \cite{busoniu2011approximate}.

\vspace{2mm}
\subsubsection{Categories of RL algorithms}

Most RL algorithms can be classified into being either \emph{model-free} or \emph{model-based} \cite{sutton1998reinforcement}. A model intuitively is an abstraction that an agent can use to predict how the environment will respond to its actions: i.e., given a state and the action performed therein by the agent, a model can predict the (expected) resultant next state and the accompanying reward. We will be mostly interested in stochastic models which can predict probabilistically possible next states and rewards given the current state and action.

In the \emph{model-based approach}, the agent builds a model of the environment through interaction with it typically in the form of a MDP analogous to the approach taken in adaptive control \cite{kumar1985survey}. With a model in hand, given a state and action, the resultant next state and next reward can be predicted allowing \emph{planning} through which a future course of action can be contemplated by considering possible future situations before they are actually experienced. Based on the MDP model in the model-based approach, a planning problem is solved to find the optimal policy function with techniques from the related field of \emph{dynamic  programming} \cite{russell1995artificial} \cite{sutton1998reinforcement}.\footnote{The term dynamic programming was originally used in the 1940s by Richard Bellman to describe the mathematical theory of multi-stage decision processes in which one needs to make the best decision one stage after another. The term `dynamic' in `dynamic programming' refers to the temporal aspect of multi-stage decision making while `programming' refers to optimization.}  Commonly used algorithms used to solve MDPs include the celebrated dynamic programming algorithms of value iteration\cite{bellman1957} and policy iteration\cite{howarddynamic}.

In the \emph{model-free approach}, on the other hand, the agent aims to \emph{directly} determine the optimal policy by mapping environmental states to actions without constructing a MDP model of the environment. Early RL systems were explicitly trial-and- error learners and were generally devoid of planning. Popular model-free RL techniques include temporal difference (TD) learning (in which a guess is updated on the basis of another guess) and Q-learning \cite{sutton1998reinforcement}. Modern reinforcement learning spans the whole gamut of approaches from low-level, trial-and-error learning to high-level, deliberative planning \cite{sutton1998reinforcement}.

RL tasks can be also be categorized into two types depending on whether the decision making tasks are sequential or not. In \emph{non-sequential tasks}, expected immediate payoff is more important, and the objective is to learn a mapping from situations to actions that maximizes the expected immediate payoff. Such learning has been studied extensively in the field of \emph{learning automata}. In \emph{sequential tasks},  the objective now is to maximize the expected long-term payoffs. Sequential tasks are considered more difficult since the chosen action may influence future trajectory of situations and payoffs. Such learning has been the subject of fields such as \emph{dynamic programming}.

\vspace{2mm}
\subsubsection{Major reinforcement learning techniques}

It is noted that RL is best understood as a class of learning problems rather than as a fixed set of algorithms or techniques. Indeed, there is great diversity in the various approaches taken by different RL algorithms and techniques.

We can broadly categorize RL techniques into two main categories of \emph{value iteration} and \emph{policy iteration} techniques. In \emph{\emph{value iterating}} learning techniques, the optimal policy is calculated on the basis of optimal value function calculated as described in section \ref{sec:VI}. In \emph{policy iterating} learning techniques, on the other hand, the learning is directly in the policy space as described earlier in section \ref{sec:PI}. We will present representative techniques that belong to these two categories next. In particular, we will discuss \emph{Q-learning} as an example value-iterating model-free technique, and will then discuss \emph{learning automata} as an example technique that is policy-iterating.

%the \emph{Q-value} or the state-action value is estimated through the \emph{Q-function} which helps determine the utility of %performing a certain action in a given state. The Q-values after their convergence are used to determine the policy that drives %the behavior of the agent.

\vspace{2mm}
\emph{\uppercase{Q-learning:}}
\label{sec:qlearning}
\vspace{2mm}

Q-learning, proposed by Watkins in 1992 \cite{watkins1992q}, is a popular value-iteration model-free technique with limited computational requirements that enables agents to learn how to act optimally in controlled Markovian domains. The implication of being model-free is that Q-learning does not explicitly model the reward transition probabilities of the underlying process. Q-learning proceeds instead by estimating the value of an action by compiled over experienced outcomes using an idea known as \emph{temporal-difference (TD) learning}.

The TD learning idea has been referred to as the central key idea in the theory of RL. TD learning combines ideas from \emph{Monte Carlo (MC) methods} and dynamic programming (DP). Like MC methods, TD method is a simulation based model-free method that can learn directly from raw experience  without a model of the environment's dynamics. Like dynamic programming, TD method used bootstrapping to update estimates based in part on other learned estimates. The concepts of TD, DP and MC are central recurring themes in RL literature.

Q-learning proceeds by incrementally improving its evaluations of the \emph{Q-values} that incorporate the quality of particular actions at particular states. The evaluation of the action-value pair, or the Q-value, is done by learning the \emph{Q-function} that gives the expected utility of taking a given action in a given state and following the optimal policy thereafter. The Q-function is defined as follows:

\begin{equation}
Q(s,a) = \sum_{s'} P_a(s,s') ( R_a(s,s') + \gamma V(s') )\
\end{equation}

The array $Q$ is updated directly with experience in the following way. The core of the update algorithm below is based on value iteration (discussed earlier in section \ref{sec:VI}). $R_{t+1}$ is the reward observed after performing $a_{t}$ in $s_{t}$, and where $\alpha_t(s, a)$ ($0 < \alpha \le 1$) is the learning rate (may be the same for all pairs). The discount factor $\gamma$ $0 \le \gamma \le 1$) trades off the importance of sooner versus later rewards. The Q-function estimate is refined in every learning step and a new policy is generated on its basis which drives the next action to execute.

%\begin{eqnarray*}
%Q_{t+1}(s_{t},a_{t}) = \underbrace{(1-\alpha_t(s_t,a_t))}_{\rm inverse~learning~rate} \times \underbrace{Q_t(s_t,a_t)}_{\rm old~value} + \underbrace{\alpha_t(s_t,a_t)}_{\rm learning~rate} \\
%\times \left[ \overbrace{\underbrace{R_{t+1}}_{\rm reward} + \underbrace{\gamma}_{\rm discount~factor} \underbrace{\max_{a}Q_t(s_{t+1}, a)}_{\rm estimate~of~optimal~future~value}}^{\rm learned~value}\right]
%\end{eqnarray*}

\begin{align}
Q_{t+1}(s_{t},a_{t}) = \underbrace{(1-\alpha_t(s_t,a_t))}_{\rm inverse~learning~rate} \times \underbrace{Q_t(s_t,a_t)}_{\rm old~value} \notag\\
+ \underbrace{\alpha_t(s_t,a_t)}_{\rm learning~rate}
\times \underbrace{(R_{t+1} + \gamma \max_{a}Q_t(s_{t+1}, a))}_{\rm learned~value}
\end{align}

Q-learning in its simplest setting stores data in tables. This quickly becomes impractical for complex systems.  In such cases, Q-learning can be combined with function approximation: in particular, (adapted) ANNs have been proposed for function approximation for large-scale RL problems \cite{tesauro2002programming}.

Q-learning does not systematically handle the tradeoff between exploration and exploitation, relying instead on heuristic explorations. Fortunately, it has been shown that Q-learning does eventually find the optimal value of an action (the proof relies on infinitely many observations for every action and state \cite{watkins1992q}). The Markovian environment of MDPs is crucial for guaranteed convergence, and the convergence guarantee is lost if this assumption is not valid.

In its basic setting, Q-learning is intended for single-agent environments, although \emph{multi-agent Q-learning}, also known as \emph{Q-learning with games}, have also been proposed recently. Multi-agent learning is especially challenging since it operates in non-Markovian environments (as the output of an action no longer only depends on the current state and agent's personal action). As such, the convergence guarantees of MDP do no extend to multi-agent RL environments due to their non-Markovian nature.

\vspace{2mm}
\emph{Application of Q-learning to routing and CRNs}: Boyan et al. showed in 1994 that routing packets through a communication network is a natural application for RL algorithms \cite{boyan1994packet}. Their `Q-routing' algorithm learned a routing policy that minimizes total delivery time by learning through experimentation with different routing policies. The presented RL based algorithm had the desirable features that: \emph{i)} its learning is continual and online, \emph{ii)} it uses local information only, and \emph{iii)} it is robust in the face of dynamic network conditions. This early paper showed that adaptive routing is a natural domain for reinforcement learning. Q-learning is perhaps the most popular model-free reinforcement learning technique which has been applied to CRNs extensively \cite{bkassinysurvey}. We refer the interested reader to a survey paper for more details and references \cite{al2013application}.

\vspace{4mm}
\emph{\uppercase{Learning Automata:}}
\vspace{2mm}

Learning automata (LA) is an AI technique that subscribes to the policy iteration paradigm of RL \cite{nicopolitidis2011adaptive} \cite{akbari2010mobility} \cite{akbari2010intelligent}. In contrast to other RL techniques, policy iterators operate by directly manipulating the policy $\pi$. Another example of policy iterators are evolutionary algorithms.\footnote{We shall discuss in section \ref{sec:evolutionary} how evolutionary algorithms share certain attributes with RL: e.g., both depend on exploration and exploitation.}

A learning automaton is a finite state machine that interacts with a stochastic environment and attempts to learn the optimal action (that has the maximum probability to be rewarded) offered by the environment so that it can ultimately choose this action more frequently than other actions. Since wireless networks operate in dynamic time-varying environments with possibly unknown characteristics (e.g., variable link qualities, dynamic topologies, changing traffic patterns, etc.), the application of LA techniques for building adaptive protocols in such networks is particularly appealing. In this regard, LA has been used in the design of wireless MAC, routing and transport-layer protocols \cite{nicopolitidis2011adaptive}.

We will now present some \emph{example LA based routing protocols}. Torkestani et al. have proposed using LA for multicast routing in mobile ad-hoc networks or MANETs\footnote{MANETs share an important characteristic with CRNs in that both of them have highly dynamic topology. The dynamically changing topology in MANETs is due to node mobility while in CRNs it is due PU arrivals.} to find routes with expected higher lifetimes through prediction of node mobility \cite{akbari2010mobility}. Another LA-based distributed broadcast solutions can be seen at \cite{akbari2010intelligent}.

\vspace{2mm}
\subsubsection{Central issues in reinforcement learning}
\label{sec:centralissuesinRL}

Some pressing issues in RL research have been highlighted \cite{kaelbling1996reinforcement} to be: trading off exploration and exploitation, learning from delayed reinforcement, making use of generalization, dealing with multiple agent reinforcement learning, constructing empirical models to accelerate learning, and coping with hidden state. Out of these issues, the issues of exploration and exploitation and that of multi-agent reinforcement learning are most relevant to our work, and we discuss them next.

\vspace{2mm}
\emph{Issue of exploration and exploitation:} Exploitation would entail favoring immediate payoff while exploration would require tolerating momentary \emph{regret} of not using the best currently known policy for the opportunity of potential information about better policies. It should be apparent after some reflection that neither exploration nor exploitation can be pursued exclusively without failing at the task of selection of the optimal action. The tension between exploitation and exploration is typified in the so-called \emph{multi-armed bandit problems}. The $k$-armed bandit problem is the simplest possible RL problem \cite{kaelbling1996reinforcement} and represent an MDP with a single state (see figure \ref{fig:markovrelation}) in which $k$ actions are available. The problem is called a $k$-armed bandit in a metaphorical reference to predicament of a gambler who must select from $k$ slot machines, colloquially called a 1-armed bandit, in a casino. Interestingly, the conflict between delayed versus immediate gratification is a dilemma unique not only to RL, the conflict it arises can be experienced in our own humanness.\footnote{It has been said by a mathematician Peter Whittle that  ``bandit problems embody in essential form a conflict evident in all human action: information versus immediate payoff.''.} Fortunately, a method has been devised by Gittins in 1979 for optimally solving the exploration and exploitation tradeoff for the simple case of k-armed bandit problem \cite{gittinsmulti} assuming a discounted expected reward criterion. This method entails providing a dynamic `\emph{allocation index}' to each action for each step in k-armed bandit problems. Gittins showed that it is guaranteed that choosing the action with the largest index value will lead to optimal balance between exploration and exploitation \cite{gittinsmulti}.

For the general case of MDPs, the optimal balance between exploration and exploitation is known to be an intractable problem to solve \cite{sutton1998reinforcement}. Therefore, a lot of interest has focused on development of heuristic or approximate methods to handle the tradeoff between exploration or exploitation. To manage the exploration or exploitation dilemma, the \emph{$\epsilon$-greedy} strategy is to select the greedy action (one that exploits prior knowledge and provides the best value) all but $\epsilon$ of the time, and to select an action randomly for the remaining $\epsilon$ of the time. The value $\epsilon$ ranges between 0 and 1 and it is possible to change this value over time. Intuitively, it would be prudent for an agent to be more of an explorer initially (by having a higher $\epsilon$) since it has no knowledge to exploit it. With passing time, as good states and actions are learnt, the agent can benefit more by being an exploiter and taking the greedy approach  (with smaller $\epsilon$) which chooses good actions more often. It makes intuitive sense that during explorations, the choice of actions are not completely random but based on some estimation of their potential value. In this regard, a \emph{soft-max action selection} technique can be used which uses the \emph{Gibbs or Boltzmann distribution} for selecting the action to explore where the probability of selecting an action is proportional to its perceived value (e.g., its Q-value). We note here the question of exploration vs. exploitation is central not only to reinforcement learning, but also to genetic algorithms, and to evolutionary algorithms in general \cite{vcrepinvsek2013exploration}.

\vspace{2mm}
\emph{Multi-agent Reinforcement Learning (MARL)}: MARL are more challenging than single-agent RL problems mainly since the Markov property does not hold in such environments as an agent's reinforcement depends not only on its current state but also on the action taken by the other agents. Accordingly, convergence guarantees that apply to MDP RL tasks do not extend in such non-Markovian MARL settings. Learning automata based tools have been quite popular in MARL environments. A detailed survey of multi-agent reinforcement learning algorithms is presented in \cite{busoniu2008comprehensive}.

\vspace{2mm}
\subsubsection{Application of RL to routing and CRNs}

The authors of \cite{di2010learning} present the benefits and potential drawbacks of using RL in CRNs. The main benefits listed are adaptivity, network awareness, and ease of distributed implementation, while the main drawback is slow convergence (although, it is pointed that convergence is not a main goal in CRNs since the environment is not stationary in any case). This paper \cite{di2010learning} also surveys the existing RL schemes in the context of ad-hoc CRNs, and proposes modifications from the viewpoint of routing and link-layer spectrum-aware operations. Another survey paper \cite{al2013application} presents a detailed survey of applications of reinforcement learning to routing in distributed wireless networks is presented in \cite{al2013application}. The interested readers are referred to these papers, and the references therein, for a more exhaustive treatment of RL applications for routing in wireless networks in general. Applications of reinforcement learning to CRNs in general are explored in \cite{yau2010applications} while RL techniques for context awareness and intelligence in wireless networks are reviewed in \cite{yau2012reinforcement}. Since CRs often have to work in unknown environments, RL seems be a promising solution to the various learning problems in CRNs and it looks set to become a popular tool for future CRN designers.

\subsection{\uppercase{Learning with Game theory}}

While game theory is essentially concerned with the decisions made by individuals in their interactions with other decision makers and their environment, researchers have long recognized the need to guide future decisions from the history of past experience. There is a lot of work on the important relationship between game theory and learning \cite{fudenberg1998theory}. A branch of game theory known as `learning game theory' studies the dynamics of individuals who repeatedly play a game, and adjust their behavior over time as a result of their experience (through, e.g., reinforcement, imitation, or belief updating) \cite{izquierdo2012learning}.

It is worth highlighting the work that has been done in identifying the similarities between inference and learning in the fields of machine learning and game theory \cite{rezek2008similarities}. In the field of game theory, learning is used implied to mean inference of the correct strategy to play against an opponent within a dynamic game (repeated game, stochastic game, or evolutionary game). Some of the models that have been used for learning in game theory include reinforcement learning, learning by imitation, myopic response, fictitious play, and rational learning \cite{izquierdo2012learning}. As examples, we discuss fictitious play, and Q-learning with game theory.

\emph{Fictitious play:} The main idea in fictitious play is that each player would choose their best strategy in each period, based on the predicted strategy that each opponent player would choose in that period, to maximize expected payoff.

\emph{Q-learning with game-theory:} Although Q-learning in its basic form is used in a single-agent RL setting, it has been extended to produce the Nash Q-learning algorithm for multi-agent RL setting based on the concept of stochastic games \cite{hu2003nash}. In this multi-agent Q-learning algorithm, the Q-value is updated with the future payoff so that each agent can observe and estimate the payoff for using a particular strategy (not only for itself but also for the other players).

A detailed survey of strategic learning in CRNs, and various spectrum access games, is presented in \cite{van2009spectrum}.

\subsection{\uppercase{Online Learning Algorithms}}

\emph{Online learning algorithms} address the task of online sequential decision-making under partial information. An example problem which can be addressed through online learning techniques is determining what route to use to drive to work everyday in an uncertain environment where the congestion pattern on the various paths is both stochastic and unknown \cite{blum2007learning}. The basic setting is we have a space of $N$ actions, from which the algorithm chooses an action (in our example, selecting the route to take) one time step after the other. The environment then makes its `move' (in our example, by setting the path congestions for that time step). The algorithm then incurs the \emph{`loss'} for its action chosen (in our example, this is how long the route took). Online learning algorithms aim to perform well in such tasks of repeated decision making. While this example (which is from \cite{blum2007learning}) relates to routing in a transportation network, it is analogous and directly extensible to the problem of routing in a CRN.

A key technique for analyzing the performance of online learning algorithms is \emph{regret analysis}. This captures our sentiment that we want our sophisticated online algorithm (which may be choosing different actions at different times) to be at least as good as some simple fixed alternative policy $\lambda$ that sticks with just one action at the time of all decisions---this will minimize our \emph{regret} of not choosing the alternative policy $\lambda$. More formally, this regret is defined to be the difference between the loss of our learning algorithm and the loss using the alternative policy $\lambda$. This regret is more properly called \emph{external regret} when the alternative policy is a \emph{static} policy (i.e., a policy of performing the same action in all time steps). External regret allows us a general methodology for developing online algorithms whose performance is comparable to that of an optimal static online algorithm. Stronger notions of regret include \emph{internal} or \emph{swap regret} which allow comparison of online action sequences in which every occurrence of a given action $i$ is changed by an alternative action $j$ \cite{blum2007learning}. There has been a lot of work by the learning theory and the game theory communities in this area, and online learning algorithms have been shown to have strong performance guarantees \cite{blum2007learning} with decision-making algorithms (such as the \emph{weighted majority algorithm}  \cite{littlestone1989weighted}) available that approach zero regret even against a fully adaptive adversary.

\vspace{2mm}
\subsubsection{Online learning algorithms in CRNs}

Han et al. proposed a using the solution concept of correlated equilibrium for opportunistic spectrum access in CRNs using a distributed no-regret learning algorithm. It was shown in their work that their correlated equilibrium based solution returns fairer results with better performance \cite{han2007distributive}.

\vspace{2mm}
\subsubsection{Online learning algorithms for routing}

Awerbuch et al. have formulated the problem of determining a sequence of routing paths in a network with unknown link delays varying unpredictably over time \cite{awerbuch2008online} as a generalization of the online multi-armed bandit problem. The sequential decision-making under partial information in this multi-armed bandit problem is handled through the framework of a repeated game with two players (algorithm and adversary) interacting over time. They have proposed two randomized online algorithms as a solution to this problem. Avramopoulos et al. have proposed using online learning algorithms as a framework for adding adaptivity to routing decisions in realistic Internet-like environments \cite{avramopoulos2008optimization}. In another work, Bhorkar et al. have presented a no-regret routing algorithm for wireless ad-hoc networks \cite{bhorkar2010no}.

\subsection{\uppercase{Neural Networks}}

Artificial neural networks (ANN or simply NN) are composed of artificial `neurons' interconnected together in a programming structure that aims to mimic the neural processing (organization and learning) of biological neurons and its behavior \cite{tsagkaris2008neural}. More specifically, NNs involve a network of simple elements that can exhibit complex global behavior determined through: \emph{i)} the way these elements are connected together into a network, and \emph{ii)} the adaptive element parameters which are tuned by a learning algorithm. ANNs are mostly used in supervised learning settings but can also be used in reinforcement learning environments (e.g., it can be used along with dynamic programming \cite{bertsekas1995neuro}, in what is known as neuro-dynamic programming, to solve RL problems) and in unsupervised learning environments (e.g., a self-organizing map (SOM) is a type of ANN that works under the unsupervised learning paradigm to produce a low-dimensional map of the input space of the training samples, called a map).

NNs are essentially ``a network of weighted, additive values with nonlinear transfer functions'' although its coined name seems to elicit a grander impression\footnote{It has been claimed that the selection of the name ``neural network'' was one of the great PR successes of the twentieth century since it sounds much more exciting by eliciting a comparison with an actual neural network (i.e., the brain) \cite{NN-name}.}.

The simplest kind of NN is a single-layer \emph{perceptron} network which is a simple kind of a feed-forward network (i.e., a network in which connections between the units do not form a directed cycle). In such a network, there exists a single layer of output nodes which is provided the input directly via a series of weights. The sum of the weighted input is calculated at each node to calculate an overall value which is then matched against a threshold (typically 0). If the calculated value is more than the threshold, the neuron is fired and it takes an activated value (typically 1), otherwise, the neuron takes a deactivated value (typically -1). Despite having a simple and efficient learning algorithm, single-layer networks are of limited utility since they have limited expressive power (i.e., they can not express complex functions) and can only learn linear decision boundaries in the input. Multi-layer networks, on the other hand, are much more expressive and can represent non-linear functions. In multi-layer NNs, processing elements are arranged in multiple layers (typically interconnected in a feed-forward fashion) with each neuron in a layer having directed connections to the neurons of the subsequent layer. Such networks have a downside that they are hard to train because of high dimensionality of the weight-space and the abundance of local minima \cite{russell1995artificial}.

\begin{table*}
\caption{Relationship between some of the fields whose techniques are presented in this paper}
\centering
\begin{tabular}{l p{4.5cm} p{4.5cm} p{3.5cm}}
\toprule
 &  \textbf{\emph{Optimal Control (OC)}} &  \textbf{\emph{Genetic Algorithms}}  &  \textbf{\emph{Game Theory}}\\
\midrule
  \textbf{\emph{Game Theory (GT)}} & Multiplayer competitive OC process \cite{haykin2005cognitive}; Dynamic Games \cite{basar1995dynamic}&  Field of Evolutionary Game Theory; Evolutionary Algorithms and GT  & -
  \\
  \\
  \textbf{\emph{Reinforcement Learning (RL)}} & Direct Adaptive Optimal Control \cite{sutton1992reinforcement};
  Adaptive Dynamic Programming &  `Exploration and Exploitation' concept;
    Evolutionary algorithms for RL \cite{kaelbling1996reinforcement}& Multiagent RL; Markov games \cite{littman1994markov}
  \\
  \\
  \textbf{\emph{Markov Decision Process}} & OC problem with a defined model; Dynamic~Programming \cite{bertsekas1995neuro}&  `Exploration and Exploitation' concept  & Stochastic or Markov games \cite{haykin2005cognitive} \cite{littman1994markov}\\
 \bottomrule
\end{tabular}
\label{tab:connections}
\end{table*}

NN is essentially a black-box statistical modeling technique that does not utilize the domain's subject knowledge but learns feature from the data itself. Despite the black-box modeling style of NN, it is a remarkably versatile tool and applies to a wide range of problems and performs fairly well in general. This has led to John Denker to famously remark that ``neural networks are the second best way of doing just about anything.'' \cite{russell1995artificial}. Notwithstanding this claim, for certain types of tasks (e.g., pattern recognition, speech recognition, etc.), NN is arguably the most effective learning method known currently \cite{mitchell1997machine}. The price of the generality on NNs, though, can be the need of large amounts of training data and in its greater convergence time.

\vspace{2mm}
\emph{Application of NNs to CRNs}:
NNs have been successfully applied to various problems in CRNs such as spectrum sensing, spectrum prediction \cite{tumuluru2010neural}, and dynamic channel selection \cite{baldo2009neural}---these last two applications are especially relevant to our focused topic of routing in CRNs. NNs have been directly employed for the problem of routing in \cite{ju2010scalable} and \cite{barbancho2006sir}. For more details about application of NN to CRNs, the interested reader is referred to the following survey papers: \cite{he2010survey}\cite{bkassinysurvey}\cite{tsagkaris2008neural}.

\subsection{\uppercase{Evolutionary Algorithms}}
\label{sec:evolutionary}

Evolutionary algorithms are a set of machine learning techniques that aim to imitate the robust procedures and structures that various biological organisms have used for adaptation and learning in their evolution. Evolutionary algorithms are similar to reinforcement learning algorithms in that they also depend on exploration and exploitation \cite{vcrepinvsek2013exploration}.

\vspace{2mm}
\emph{\uppercase{Genetic algorithms:}} A genetic algorithm (GA) is a particular class of evolutionary algorithm which uses techniques such as inheritance and natural selection which are inspired from evolutionary biology \cite{goldberg1988genetic}. In particular, GA fundamentally relies on the genetic operators of random \emph{mutation} and recombination through \emph{crossover} to improve the current solution. Apart from these operators, the design of GAs also includes other crucial components such as population initialization, genetic representation, fitness function, and a mechanism for selection.

Genetic algorithms are typically implemented in a computer simulation in which evolutionary techniques (mutation, crossover, etc.) are applied on a \emph{population} of candidate solutions (called \emph{individuals}). Individuals are encoded in an abstract representation known as a \emph{chromosome} (which may be problem specific although representation in strings of 1s and 0s is common). The evolution can start from a population of completely random individuals and can evolve to better solutions through \emph{survival of the fittest} after application of genetic operators in every \emph{generation}. In every generation, multiple individuals are stochastically selected from the current population with fitter individuals more likely selections and are genetically modified (mutated or recombined) to form the next generation of the population. The usage of genetic operators and stochastic selection allow a gradual improvement in the `fitness' of the solution and allow GAs to keep away from local optima.

Like neural networks, genetic algorithms apply very generally. John Denker's quote about NNs that ``neural networks are the second best way of doing just about anything'' can be supplemented with the addition ~``... and genetic algorithms are the third best.'' \cite{russell1995artificial}. Neural networks and genetic algorithms can be thought of as the sledgehammers of the algorithms craft due to their broad applicability and can be readily invoked when more specialized methods fail. Predictably, this generality can come sometimes at a cost in performance and time to convergence. However, these tools are worth an initial try and may perform very well for certain problems. Therefore, Denker's remark must be construed to corroborate the observation that GA are widely applicable, but it may undersell some desirable features of GA (and NN) which can make it an ideal tool for certain problems.

\begin{table*}
\caption{Summary of the various learning techniques discussed in section \ref{sec:learning}}
\centering
\begin{tabular}{l p{7cm} p{7cm}}
\toprule
 \textbf{\emph{Learning techniques}} &  \textbf{\emph{Application to CRNs}} & \textbf{\emph{Applications to routing}}\\
\midrule
%\vspace{2mm}
\textbf{\emph{Hidden Markov model}} & Spectrum occupancy prediction: \cite{akbar2007dynamic} \cite{park2007hmm} \cite{choi2013estimation}; Spectrum sensing, primary signal detection (see references in \cite{he2010survey})  & Can indirectly utilize spectrum occupancy and channel quality predictions\\
%\vspace{2mm}
\textbf{\emph{Reinforcement learning}} &  Many applications: see survey paper \cite{bkassinysurvey} & Q-routing algorithm \cite{boyan1994packet}; Learning automata \cite{akbari2010intelligent} \cite{akbari2010mobility}; see further references in survey papers \cite{bkassinysurvey} \cite{al2013application} \cite{yau2010applications}. \\
%\vspace{2mm}
\textbf{\emph{Learning with games}}  & Spectrum access games \cite{van2009spectrum}  & Various routing games: see \cite{pavlidou2008game}  \\
%\vspace{2mm}
\textbf{\emph{Online learning}}  & Opportunistic spectrum access \cite{han2007distributive}  & No regret routing for adhoc networks \cite{bhorkar2010no}  \\
%\vspace{2mm}
\textbf{\emph{Genetic algorithms}} & Modeling wireless channel: \cite{rondeau2004online}    & Shortest path routing \cite{ahn2002genetic} \\
%\vspace{2mm}
\textbf{\emph{Swarm algorithms}}  & Adaptive optimization \cite{mahdi2012adaptive}    & Shortest path routing \cite{mohemmed2008solving}   \\
%\vspace{2mm}
\textbf{\emph{Ant colony optimization}}  & Cognitive engine design  \cite{zhao2012cognitive}    & Routing with ACO in CRNs \cite{zhao2012cognitive} and MANETs \cite{di2005anthocnet} \\
%\vspace{2mm}
\textbf{\emph{Neural networks}}   & Spectrum occupancy prediction \cite{tumuluru2010neural}; Dynamic channel selection \cite{baldo2009neural}; Radio parameter adaptation (see references in \cite{he2010survey}). & Routing with NNs \cite{ju2010scalable} and \cite{barbancho2006sir} \\
%\vspace{2mm}
\textbf{\emph{Support vector machines}}  & Spectrum sensing, signal classification and pattern recognition (see references in \cite{bkassinysurvey})   & - \\
%\vspace{2mm}
\textbf{\emph{Bayesian learning}}  & Establishing PU's activity pattern \cite{saad2012cooperative} \cite{han2011repeated}; Channel estimation \cite{haykin2005cognitive}; Channel quality prediction  \cite{xing2013spectrum}  & Bayesian routing in delay-tolerant-networks \cite{ahmed2010bayesian} \\
 \bottomrule
\end{tabular}
\label{tab:learning}
\end{table*}

We have noted earlier that evolutionary algorithms, and by extension GA, are related to reinforcement learning in that both depend on exploration and exploitation \cite{vcrepinvsek2013exploration}. Evolutionary algorithms based learning also illustrates how learning can be viewed as a special case of optimization. These algorithms pursue the `optimization problem' of finding the optimal hypothesis according to a predefined fitness function \cite{mitchell1997machine}. With the insight that learning is ultimately related to optimization, we can apply other optimization and heuristic techniques to machine learning problems. For a discussion about heuristic optimization techniques (such as simulated annealing, tabu search, hill climbing) and their application to CRNs, readers are referred to \cite{he2010survey}.

A recurrent theme in this paper is that most of the machine learning fields, and techniques from kindred disciplines, are more closely related than immediately apparent. Some representative connections between various fields discussed in this paper (such as game theory, MDPs, RL, GA and optimal control) are tabulated in table \ref{tab:connections} for easy reference.

\vspace{2mm}
\emph{Application of GAs to Routing}: Ahn et al. proposed tackling the shortest path routing problem through GAs \cite{ahn2002genetic}. The paper discussed the issues of  path-oriented encoding, and path-based crossover and mutation, which are relevant to the issue of routing \cite{ahn2002genetic}.

\vspace{2mm}
\emph{Application of GAs to CRNs}: An early application of GA techniques to CRNs is documented in a paper authored by Rondeau et al. \cite{rondeau2004cognitive}. This paper presented the adaptation mechanism of a cognitive engine implemented by the authors which used GAs to evolve a radio's parameters to a set of parameters that optimize the radio for the user's current needs. This paper also proposed a GA approach, called the `wireless system genetic algorithm' (WSGA), to realize cross-layer optimization and adaptive waveform control \cite{rondeau2004cognitive}.

\vspace{2mm}
\emph{\uppercase{Swarm intelligence:}} Swarm intelligence refers to a class of machine learning techniques in which it is aimed that intelligence shown in social cooperative animals (such as ants and bees) be replicated in a distributed \emph{computational} setting. Such animals are well known to form communities that display emergent behavior as the simple limited individuals collaborate to display complex intelligent behavior. Swarm intelligence techniques in general emphasize distributed implementation and coordination through communication. Solutions based on swarm intelligence techniques have been proposed both for CRNs  \cite{mahdi2012adaptive} and for routing problems \cite{mohemmed2008solving}.

\vspace{2mm}
\emph{\uppercase{Ant Colony Optimization:}} While typical `shortest path' routing protocols may have significant computational and message complexity, the humble biological ants, in a marvel of nature, are able to shortest routes to food sources in the dynamics of ant colony with extremely modest resources. A lot of research effort has been focused on imitating the performance of biological ants to produce optimized and efficient distributed routing behavior \cite{zhao2012cognitive} \cite{di2005anthocnet}.

\subsection{\uppercase{Support Vector Machine}}

Support vector machines (SVM) is a supervised learning technique used mainly for tasks such as pattern recognition and classification. SVMs belong to the general family of learning methods known as \emph{kernel machines} \cite{russell1995artificial}. SVMs can both \emph{i)} use an efficient training algorithm and, \emph{ii)} represent complex, non-linear functions. SVMs typically outperform ANNs for small training examples, but require prior knowledge of the observed process' distribution and labeled data \cite{bkassinysurvey}.

\vspace{2mm}
\emph{Application of SVMs to CRNs}: SVMs have been applied to CRNs, but its application has been mostly limited to problems of signal classification \cite{bkassinysurvey}. Due to its supervised learning style, it does not seem like SVMs will have a direct role to play in the design of AI-based routing protocols for CRNs.

\subsection{\uppercase{Bayesian Inference}}

Bayesian analysis accords significant importance to the \emph{prior distribution} which is supposed to represent knowledge about unknown parameters before the data becomes available.  While it is a common assumption that the agent has no prior knowledge about what it is trying to learn, this is not an accurate reflection of reality in many cases. Frequently, an agent will have  some prior information, and the learning process should ideally exploit this available information.

Bayesian learning can be viewed as a form of uncertain reasoning from observations \cite{russell1995artificial}. Bayesian learning is used to calculate the probability of each hypothesis, given the data, and to make predictions on that basis. It has been shown that the true hypothesis eventually dominates in Bayesian prediction \cite{russell1995artificial}.

Bayesian analysis is appealing since it provides a mathematical formulation of how previous knowledge can be incorporated with fresh evidence to create new knowledge. However, choosing the right prior distribution is not trivial and an incorrect assumption can skew the inference. It is for this reason that some statisticians feel uneasy about the use of prior distributions fearing that it may distort ``what the data are trying to say.'' \cite{box2011bayesian}. We can model the prior distribution to prior knowledge or use a `noninformative' prior to model ignorance about prior information.

\emph{Bayesian networks} can be used for computing how much a set of mutually exclusive prior events contributes to a posterior condition, which can be a prior to yet another posterior, and so on. Bayesian networks can be used for reasoning and for tracing chain of conditional causation back from the final condition to the initial causes \cite{russell1995artificial}. Previous work on using Bayesian networks for reasoning in CRNs has been summarized in \cite{adamopoulou2008enhancing}.

While it was noted by Rondeau in 2009 \cite{rondeau2009artificial} that little research attention has focused on using Bayesian methods of statistical inference in CRNs, a lot of Bayesian inference based work have recently been proposed for CRNs. In particular, \emph{Bayesian non-parametric models} have been applied to CRNs by various researchers \cite{saad2012cooperative} \cite{han2011repeated} due to their desirable characteristics such as its ability to flexibly model an unknown environment with model complexity growing as warranted by new data. Parametric models typically assume some finite set of parameters $\theta$ and assume that $\theta$ captures everything there is know about the data. Non parametric models, on the other hand, do not assume that the data distribution can be explained on the basis of a finite set of parameters---instead, an infinite dimensional set of parameters $\theta$, envisioned as a function, is assumed. Bayesian non-parametric models typically exploit in their formulation decades of research on Gaussian processes (which defines a distribution on functions) and Dirichlet process (which defines a distribution on distributions). Popular Bayesian nonparametric techniques that use these processes include Gaussian process regression, in which the correlation structure is improved as the sample size increases, and Dirichlet process mixture models for clustering, which adapt the number of clusters to the complexity of the data.

\vspace{2mm}
\emph{Applications of Bayesian non-parametric methods in CRNs}: Saad et al. have proposed a cooperative Bayesian nonparametric framework for primary user activity monitoring in CRNs \cite{saad2012cooperative}. In another work, spectrum access in CRNs was modeled as a repeated auction game and a Bayesian nonparametric belief update scheme was constructed based on the Dirichlet process \cite{han2011repeated}.

\vspace{2mm}
\emph{To conclude this section on learning algorithms, a representative summary of this section on learning techniques for CRNs is captured in table \ref{tab:learning}.}

\begin{table*}
\caption{Summary of representative routing protocols that have been proposed for CRNs}
\centering
\begin{tabular}{p{2.7cm} p{1.2cm} p{3.5cm} p{8.3cm} }
\toprule
 \textbf{\textbf{\emph{Reference}}} & \textbf{\emph{Type}} & \textbf{\emph{PU model}} & \textbf{\emph{Comments}}\\
\midrule

\multicolumn{2}{l}{\textbf{\emph{Throughput maximizing}}} \\
Cacciapuoti et al. \cite{cacciapuoti2012reactive} & Reactive & Markov on-off process & Reactive routing for mobile
ad-hoc CRNs \\
Ding et al. \cite{ding2009rosa} && Not described & Cross-layer routing and dynamic spectrum
allocation algorithm \\
SAMER \cite{pefkianakis2008samer}  & Reactive & Bernoulli trial every $t$& Routes with highest spectrum availability (``least-used spectrum first'')  \\
SPEAR \cite{sampath2008high} & Reactive & Not described & Joint spectrum and route discovery with distributed path reservations to minimize inter- and intra-flow interference.  \\

\multicolumn{2}{l}{\textbf{\emph{Delay minimizing}}} \\
How et al. \cite{how2011routing} & Reactive & 2-state Semi Markov Model & Multi-metric (delay and stability) routing providing differentiated service\\
SEARCH  \cite{chowdhury2009search} & Reactive & Not described & Designed for
mobile CRNs based on geographic forwarding principles \\
CRP  \cite{chowdhury2011crp} & Reactive & Markov on-off process & Distributed joint route and spectrum selection protocol that explicitly protects PU receivers, and allows multiple classes of routes.   \\

\multicolumn{2}{l}{\textbf{\emph{Stability maximizing}}} \\
Coolest-first \cite{huang2011coolest}  & Reactive & Markov on-off process & Proposed new routing metrics to capture the time-varying effects of spectrum availability  \\
%Deng et al. \cite{deng2007collaborative} &&  Not described &  Proposes collaborative strategy for route
%and spectrum selection \\
Tuggle \cite{tuggle2010cognitive}& \emph{Proactive} & Not considered & Proposes proactive multi-path routing  \\
Gymkhana \cite{abbagnale2010gymkhana}& Reactive & Markov on-off process & Path connectivity based distributed protocol that avoids poorly connected zones\\

\multicolumn{2}{l}{\textbf{\emph{Maintenance minimizing}}} \\
Zhu et al. \cite{zhu2008stod} & Hybrid & Not described & Combines proactive routing and on-demand route discovery  \\
Filippini et al. \cite{filippini2009minimum} && Ergodic random binary process & Optimal centralized, along with, distributed algorithms proposed both for exactly and statistically known PU activity. \\

 \bottomrule
\end{tabular}
\label{tab:routing}
\end{table*}

\section{\uppercase{Routing in CRN}s}
\label{sec:routing}

\subsection{\uppercase{Traditional wireless routing protocols:}}

While our focus is on surveying techniques useful for cognitive routing protocols in the context of CRNs, it is also prudent to exploit and leverage the huge amount of previous work on routing protocols for wireless networks in general. While wireless networks include both wireless LANs and multi-hop wireless networks, our focus is going to be dominantly on multi-hop wireless networks such as mobile ad-hoc networks, wireless mesh networks and CRNs. We focus on these networks to build upon the insights that we can leverage for the design of effective routing protocols for CRNs.

Previous work on routing in multi-hop wireless networks can be noted for the most part for the lack of learning from environment. Most of the classical wireless routing protocols tend to use instantaneous online parameters and do not utilize environment history and learn from it to predict about links and parameters that are more likely to result in better quality routes. These protocols also do not learn about parameter history and therefore cannot prioritize higher-quality links over links of poor quality. While primitive protocols such as AODV, DSDV, and DSR have typically relied on basic metrics such as hop count or delay, other metrics were developed for wireless networks over time such as those that targeted: maximizing throughput \cite{de2005high}, minimizing interference \cite{subramanian2006interference}, load balancing \cite{raniwala2005architecture}, and choosing more reliable links \cite{de2005high}. Since metrics designed for traditional wireless networks do not sufficiently capture the time-varying spectrum availability found in CRNs, some recent works have proposed more nuanced spectrum aware routing metrics \cite{pefkianakis2008samer} \cite{huang2011coolest} \cite{zhu2008stod} \cite{filippini2009minimum}  \cite{caleffi2012opera}.

\subsection{\uppercase{Routing Protocols for CRN}s:}

Some \emph{challenges} for effective routing in CRNs have been highlighted by Akyildiz et al. \cite{akyildiz2006next}. It was highlighted that while spectrum sensing techniques and spectrum sharing solutions have received considerable attention by the CRN research community, routing remains yet an important unexplored area in CRNs. Akyildiz et al.  go on to highlight that the unique characteristics of the open spectrum phenomenon necessitate development of novel routing algorithms. Some other challenges include \emph{i)} intermittent connectivity with neighbors in DSA networks causing a highly dynamic topology, \emph{ii}) heterogeneous channels with diverse channel properties whose availability is time-varying \cite{akyildiz2006next}, and \emph{iii}) potential non-availability of common control channel. The challenges and issues related to common control channel are covered at depth in \cite{lo2011survey}. Another potential problem is the fact that CRs would typically have to work in unknown, or incompletely known, environments. With the strong assumption of availability of full spectrum knowledge, optimization based routing solutions have been proposed \cite{ma2008joint} \cite{shi2008distributed}.
Such works are applicable only where this assumption is justified: an example scenario being TV band whitespace networking where the SUs can query databases storing the spectrum map. For the more general case, solutions need to be devised that work with limited local spectrum knowledge.

A wide variety of routing protocols have been proposed for CRNs and a representative summary can be seen at table \ref{tab:routing}. These routing protocols have used a diverse set of routing metrics and objectives: e.g., \emph{throughput maximizing protocols} (\cite{cacciapuoti2012reactive} \cite{ding2009rosa} \cite{pefkianakis2008samer} \cite{sampath2008high} ), \emph{route-stability maximizing protocols} (\cite{huang2011coolest} \cite{tuggle2010cognitive}), \emph{delay minimizing protocols} (\cite{how2011routing} \cite{chowdhury2009search} \cite{chowdhury2011crp} \cite{yang2008local}), and \emph{route-maintenance minimizing protocols} (\cite{zhu2008stod} \cite{filippini2009minimum}).

The most commonly used approach in literature is to incorporate these metrics into some variant of a reactive or an on-demand routing protocol\footnote{See table \ref{tab:routing} to see the preponderance of reactive routing protocols proposed in literature.} to avoid the overhead of managing dynamic topologies proactively. With dynamic spectrum access (DSA) being envisioned as a prime application of CRNs, it is important for routing protocols for CRNs to incorporate PU traffic dynamics into its design. Some of the CRN routing protocols have conspicuously not catered to PU dynamics in their design \cite{cacciapuoti2012reactive} \cite{sampath2008high} \cite{chowdhury2009search} \cite{yang2008local} \cite{deng2007collaborative}, although more recent work \cite{ding2009rosa} \cite{pefkianakis2008samer} \cite{how2011routing} \cite{filippini2009minimum} have importantly incorporated PU awareness.

Sun et al. \cite{sunperformance} have conducted a detailed performance evaluation of three representative CRN routing protocols: SAMER \cite{pefkianakis2008samer}, Coolest Path \cite{huang2011coolest}, and CRP \cite{chowdhury2011crp} using both simulations (on the NS2 simulator) and an empirical evaluation (on a testbed of 6 node testbed based on USRP2 platform). The three protocols evaluated (SAMER \cite{pefkianakis2008samer}, Coolest Path \cite{huang2011coolest}, and CRP \cite{chowdhury2011crp} all have different design objectives. SAMER aims mainly at finding the highest throughput path while considering both the PU/ SU activities and the link quality. Coolest Path is designed to prefer paths that more stable since it prefers path with the highest spectrum availability. CRP is designed to either find a path with minimum end-to-end delay along with satisfactory PU protection, or to offer more complete protection to PU receivers at the cost of some performance degradation to SUs. It has been shown in their simulation and testbed results that SAMER provides the highest throughput under low PU activity (since SAMER aims to calculate throughput maximizing paths explicitly) and is also shown to be robust to packet loss; however, its performance under high PU activity deteriorates, particularly in the simulation results. Sun et al. also provide qualitative insights into the design of CRN routing protocols. Their findings suggests that taking link-quality and interference between SUs into account can great improve routing performance particularly under low PU activity. For high PU activity, however, path stability and path length become more important. Another important finding is that estimating spectrum availability based only on local observations cannot guarantee path stability therefore suggesting improvements can be made through cooperation.

Relatively few studies in the literature have addressed the \emph{multicast routing} problem in CRNs. Kim et al. have proposed a multicast routing protocol (COCAST) for mobile ad-hoc networks with nodes equipped with CRs \cite{kim2009cocast}. Their work aimed at improving the scalability of the traditional ODMRP multicasting protocol in an environment using CRs. In another work, Almasaeid et al. have addressed the problem of assisted multicast scheduling in cognitive wireless mesh networks \cite{almasaeidexploiting}, and have proposed two approaches for cooperative multicasting: the first depending on the assistance of multicast receivers in delivering multicast data to other receivers, while the second is network-coding based. Some other works that have addressed the joint problem of routing and channel assignment for multicast communication in multihop CRNs have also been proposed \cite{almasaeid2010demand} \cite{ren2009minimum}.

Broadcasting is a commonly used networking primitive used both in control and data traffic. The problem of \emph{broadcast routing} in CRNs is challenging as noted in \cite{akyildiz2006next}. In CRNs, channel heterogeneity of channels, intermittent connectivity, and lack of a common control channel can constrain the ability to perform effective broadcast routing \cite{akyildiz2006next}. Recently, a work has been proposed for fully distributed broadcast routing in CRNs without requiring a common control channel \cite{song2012distributed}. An adaptive channel assignment scheme that modifies the assignment to suit broadcast routing when the broadcasting traffic volume is significant is presented in \cite{mir2012unified}. Some other works that have addressed the problem of broadcasting in CRNs include \cite{htike2013broadcasting} \cite{fahad2010broadcasting}.

It is noted here that while most of the proposed routing protocols do include certain adaptive features, relatively little work has been done to integrate AI-based machine learning techniques into the routing solutions for CRNs. This is a promising new subfield ripe for future research exploration. To realize the vision of next-generation cognitive networks, it is imperative that due attention be given to this central piece of the overall cognitive architecture.

Interested readers are referred to the following survey papers on routing in CRNs and the references therein to find more information about the various routing protocols proposed for CRNs \cite{cesana2011routing} \cite{al2013routing} \cite{abdelazizasurvey}.

\section{\uppercase{Cognitive Routing Tasks in CRN}s}
\label{sec:cogroutingtasks}

As noted earlier, although CRN routing protocols do mostly incorporate spectrum-awareness into their design, future cognitive networks will require greater architectural support from fully `cognitive routing protocols' that will seamlessly incorporate AI-based techniques such as learning, planning, and reasoning in their design.

Some \emph{inference and reasoning} and \emph{modeling and prediction} cognitive tasks that future cognitive routing protocols must incorporate are described next in subsections \ref{plan-and-reason-tasks} and \ref{mod-and-pred-tasks}, respectively.

\subsection{\uppercase{Inference and Reasoning Tasks}}
\label{plan-and-reason-tasks}

Reasoning is an important aspect of CRN behavior and is necessary for cognitive behavior. Knowledge can be represented using an \emph{ontology} which provides shared vocabulary useful for modeling a domain, e.g., it can be used to model the type of objects and concepts existing in a system or domain, and their mutual relationship and properties \cite{gavrilovskalearning}. A rule based system can make use of a knowledge base and some means of inference through an \emph{inference engine}.

It is also possible to reason by analogy. This involves the transferring of knowledge from a past analogous situation to another similar present situation. Case-based reasoning (CBR) is a well-known kind of analogy making which has been exploited in CRN research \cite{he2010survey}. In case-based reasoning a database of existing cases is maintained and used to draw conclusions about new cases. The CBR reasoning method can utilize procedures like pattern matching and various statistical techniques to find which historical case to relate to the current case.

Fuzzy logic is another tool that is useful for reasoning in systems and situations having inherent uncertainty or ambiguity. Since complete environmental knowledge is difficult, or even impossible, to obtain in CRNs. Fuzzy logic is natural fit to the CRN environment where there is limited or no information about certain environment factors. Fuzzy logic based reasoning has been used commonly in CRNs \cite{erman2009fuzzy} \cite{shatila2012adaptive}.

While reasoning is an extremely important part of cognition, a comprehensive treatment of various reasoning tools and techniques useful for CRNs is outside the scope of this work. We refer the interested readers to a recent survey on learning and reasoning in CRNs for a comprehensive account of methods, techniques, issues and challenges in implementing reasoning \cite{gavrilovskalearning}.

\subsection{\uppercase{Modeling, Prediction, and Learning Tasks}}
\label{mod-and-pred-tasks}

Future cognitive routing protocols can benefit from the following tasks: \emph{i)} \emph{channel quality modeling and prediction}, \emph{ii)} \emph{PU activity modeling and prediction}, and \emph{iii)} \emph{detecting and mitigating selfish behavior}. We will discuss these in turn next under their respective headings.

\vspace{2mm}
\subsubsection{Channel quality modeling and prediction}

In \cite{rondeau2004online}, Rondeau et al. proposed using HMM to model the wireless channel online with the HMM being trained using a genetic algorithm. In \cite{akbar2007statistical}, techniques for modeling wireless network channel using Markov models are presented along with techniques for efficient estimation of Markov model parameters (including the number of states) to aid in reproducing and/or forecasting channel statistics accurately. In another work, Xing et al. have proposed to perform channel quality prediction using Bayesian inference \cite{xing2013spectrum}. Channel estimation problem has also been addressed in \cite{haykin2005cognitive} in which the use of particle filters, rooted in Bayesian estimation, were proposed as a device for tracking statistical variations in a wireless channel.

Researchers have proposed using an ANN-based cognitive engine for learning how various channel's quality status affects performance and thereby dynamically selecting a channel that improves performance. The dynamic selection of channels has an obvious implication for network-layer functionality and the routing algorithm for such networks should be able to keep up with the channel changes so that best performing routes are selected.

\vspace{2mm}
\subsubsection{Spectrum occupancy modeling}

A satisfactory model of spectrum occupancy (or, of spectrum white spaces)  should incorporate: \emph{i)} states of the channel along with their transition behavior, and \emph{ii)} the \emph{sojourn time} or the time duration the system resides in each of the states \cite{geirhofer2007cognitive}.

Since many DSA environments (e.g., contention based protocols such as IEEE 802.11) do not have a slotted structure, it is more appropriate to use a continuous-time (CT) model. A CT model that is especially relevant to DSA, and one that is popularly used for modeling spectrum occupancy, is the semi-Markov model (SMM) which generalizes the concept of CT Markov chains (CTMCs).  Although both the semi-Markov and CTMC models have the Markovian property and they describe the transition behavior in the same way, a SMM allows for specifying the occupancy periods, or the sojourn time, for each state arbitrarily. In particular, the occupancy time does need have to be necessarily exponentially distributed as must be the case for CTMCs by definition \cite{kleinrock1975queueing} \cite{ross1970applied}. Specification of a SMM therefore requires both the statistical specification of the transition behavior and of the sojourn time within each state \cite {kleinrock1975queueing} \cite{ross1970applied}.

It has been posited that for practical purposes of analyzing DSA/ CRNs, a  simple two-state semi-Markov ON-OFF model is adequate for modeling spectrum usage \cite{lopez2011overview} (table \ref{tab:routing} may be referred to see the popularity of this model).  The OFF state represents an idle channel, while the ON state indicates a busy channel not available for opportunistic access, with the length of ON and OFF periods being random variables (RVs) following some specified distribution. Such a model is also known as a stochastic duty cycle model \cite{wellens2010lessons}. The usage of this simple semi-Markov ON-OFF model is quite popular \cite {geirhofer2007cognitive},  although other more elaborate models are also available \cite{ghosh2010framework}. Geirhofer et al. showed in \cite{geirhofer2006dynamic} that such a model can be used to empirically model  the spectrum use in IEEE 802.11b WLAN-systems. It was noted that their results should also extend to other systems having  multi-access protocols  similar to CSMA/ CA.

An important aspect of using such semi-Markov models is specifying the state sojourn or stay times, and to study if successive period lengths are correlated. The simplest approach is to assume the state sojourn time is exponentially distributed and that successive stay times are not correlated. Such an approach is interesting due to its simplicity and tractability. Unfortunately, studies have shown that this simple approximation does not tally up well with empirical studies on actual systems \cite{geirhofer2007cognitive}. Nonetheless, exponential distributions is still used heuristically \cite{lee2008optimal}, although such an approach is not entirely justified statistically, since this assumption makes the model earlier to apply in practice.
Empirical studies have shown that state sojourn times typically have larger variability than suggested by the exponential distribution.  In fact, the distributions of the ON and in particular the OFF period were often found to be heavy-tailed \cite{geirhofer2007cognitive}.  These results motivated the need of simple models featuring correlated ON and OFF periods with heavy-tailed marginal distributions. In this regard, Pareto distributions have been used in literature \cite{wellens2009modelling}. Since, heavy-tailed distributions have the disadvantage of being difficult to analyze analytically \cite{feldmann1998fitting}, other approximate distributions have also been explored. In particular, the flexible yet tractable \emph{phase-type distributions}\footnote{Phase-type distributions result from a network of one or more inter-related Poisson processes. The distribution can be represented as a random variable modeling the time to absorption in a Markov process with one absorbing state. Due to its great flexibility, it can be used to model any positive valued distribution. Furthermore, efficient algorithms exist for estimating such a model's parameters \cite{thummler2005novel}.}  and Beta distributions \cite{wellens2010lessons} have been used to capture the data statistically.

Measurements have shown that lengths of vacant periods in a given frequency band can be correlated in addition to having a heavy-tailed distribution \cite{geirhofer2007cognitive}. As simple semi-Markov ON-OFF models cannot reproduce this effect, Wellens et al. proposed producing correlated sequences using an aggregation of multiple semi-Markov ON-OFF processes \cite{wellens2009modelling} similar to how certain self-similar traffic models work \cite{park2000self}.

The spectrum occupancy model should incorporate not only the temporal aspect of PU activity but also its \emph{spatial} aspect. The impact of PU activity pattern on \emph{spatial} spectrum reuse opportunities have been studied in \cite{riihijarvi2010impact}.

A lot of studies have focused on empirical modeling of spectrum usage and have proposed various models for PU traffic pattern \cite{wellens2010lessons} \cite{wellens2009modelling} \cite{wellens2009empirical}  \cite{harrold2011long} \cite{ghosh2010spectrum} \cite{hoyhtya2010classification}. For further details, interested readers are referred to the survey papers \cite{lopez2011overview} \cite{masontaspectrum} in which various statistical models proposed in literature for modeling temporal and spatial spectrum occupancy are reviewed in detail.

\vspace{2mm}
\subsubsection{PU activity modeling and prediction}

In DSA CRN networks, being the licensed incumbent user, a primary user (PU) has prioritized access to the wireless spectrum. Therefore, on the arrival of a PU, a SU must either vacate the relevant channel by switching to another channel or by terminating its connection; alternatively, the PU must reduce its transmission power to ensure that PU does not face any interference. Since the arrivals of PU are non-deterministic, and random from the point-of-view of a SU, frequent PU arrivals can lead to frequent temporal connection losses for secondary users thereby seriously impacting its performance.  However, a PU can probabilistically model the arrival process and traffic pattern of PU and avoid the channels that will be claimed by PU with a high probability. This can help reduce the temporal connection loss faced by SUs and potential interference faced by PUs due to any delays in vacation of channel by SUs.

A cognitive radio that manages to learn the behavioural patterns of a primary user by modeling it can optimize its performance by exploiting the learned model. For example, a SU can exploit information, potentially gleaned from spectrum sensing data, and select white spaces (that emerge due to absence of PUs) that tend to be longer lived at certain times of day and at certain locations. Knowing something about PU patterns can also be helpful for advanced planning when a SU has to decide the channel to switch to on the arrival of a PU \cite{doyle2009essentials}.

A number of techniques have been proposed for spectrum prediction including techniques that are: a) HMM based, b) NN based, Bayesian inference based, moving-average based, autoregressive-model based, and static-neighbor-graph based (which is able to incorporate PU mobility pattern) \cite{xing2013spectrum}.

\emph{HMMs} have been popularly used for spectrum occupancy prediction \cite{akbar2007dynamic} \cite{park2007hmm}. Akbar et al. utilized HMM models for predicting spectrum occupancy of the licensed radio bands for CRNs in their proposal of an HMM-based DSA algorithm \cite{akbar2007dynamic}. Choi et al. proposed a channel learning scheme based on HMM and also proposed a partially observable Markov decision process (POMDP) based framework for channel access to opportunistically exploit frequency channels a primary network operates on \cite{choi2011opportunistic}. Choi et al. have another follow up work on using HMM to model the traffic pattern on PUs \cite{choi2013estimation}.

Various models for traffic pattern prediction for PU are presented in \cite{li2008traffic}. Wang et al. \cite{wang2009primary} have proposed modeling the interaction between the PUs and SUs through continuous-time Markov chains (CTMC). Saad et al. have proposed using cooperation between CR devices that are observing the availability pattern of PUs, and the usage of \emph{Bayesian nonparametric techniques} to estimate PU activity pattern's distributions. Spectrum prediction has also been tackled through \emph{Neural Networks} in \cite{tumuluru2010neural}. For more details about spectrum prediction techniques, the interested readers are referred to a detailed survey on this topic \cite{xing2013spectrum} and the references therein.

\vspace{2mm}
\subsubsection{Detection of Selfish Behavior}

Network-layer behavior entails both the problems of routing and forwarding. In wireless networks, selfish behavior can manifest itself when nodes engage in unsocial behavior---i.e., they utilize the network resources but do not pay back the favor by providing necessary services to the other network nodes. For correct network behavior, it is important that such behavior be arrested.  The following papers have addressed the problems of identifying and mitigating selfish network behavior \cite{eidenbenz2005commit} \cite{wang2004truthful}. This problem has been studied through the tools provided by game theory in
\cite{roughgarden2007routing} \cite{zakiuddin2005towards}.

\section{\uppercase{Open research issues and future work}}
\label{sec:openissues}

%\subsection{\emph{Realizing the vision of `Cognitive Networks'}}
In this section, we will outline some of the major open research issues in building cognitive networks and in developing AI-enabled cognitive routing protocols. We will also discuss potential future work.

\subsection{Incorporation of user preferences and context-awareness}

While most of the CRN research focus has been on solving the engineering challenges in building the artifact of a cognitive radio network, the role of users, their preferences and context-awareness seem to unfortunately have taken a back-seat.  If a `wireless society' \cite{schwartz2012some} vision is to materialize in the not so distant future, it is imperative that researchers focus on seamless integration of user preferences, and awareness of (identity, location, time, activity-based) context into cognitive networking design.  There has been some initial work done in this area \cite{bantouna2012overview} but a lot more work needs to be done.

\subsection{Application of novel machine-learning techniques}

%\emph{Eliciting and modeling cooperative behavior:}
Game theory, reinforcement learning, neural networks, and genetic algorithms, due to their natural fit to the kind of problems faced in CRNs, are understandably the most used AI techniques in CRNs. However, as listed in this survey paper, there are various other machine-learning techniques that can be plausibly applied to tasks much more diverse than their current application. In particular, it is anticipated that Bayesian techniques will find increasing usage in CRNs. It is an open research question that which machine-learning techniques, apart from the current popular approaches, would prove to be most successful in solving problems in CRNs.

\subsection{Interworking with other modern technologies}

An initial promise of software defined radio (SDR) was seamless interworking with a plethora of technologies through software defined adaptations. The vision of CRNs has evolved from the foundations of SDRs and aims to provide users with seamless holistic experience that integrates potentially heterogeneous technologies. The interplay of cognitive radios with the software defined networking (SDN) architecture, which allows a standards based interface \cite{mckeown2008openflow} between a centralized `network controller'\footnote{The centralized SDN network controller can itself be built as a distributed system to be scalable and avoid a single point of failure.} and networking devices, should be explored. It is possible that interesting use cases will emerge that will synergize the mainly centralized operational paradigm of SDNs with the mainly distributed operational paradigm of CRs. While the emphasis of SDN architecture has been on the separation of control and data planes, it is worth exploring if a combined SDN and CR architecture can help realize the vision of having a `knowledge plane' for networks as envisioned by Clark et al. \cite{clark2003knowledge}. Also, it is worth exploring how cognitive networks may seamlessly integrate modern technologies like internet-of-things, pervasive and ubiquitous technology.

\subsection{Cross layer optimization for cognitive networks}

The overarching focus of the CRN research community to date has been on problems such as spectrum sensing, signal classification and other issues that relate to PHY and MAC layers. Relatively less attention has been paid to problems on the networking and higher layers. To realize the vision of cognitive networks, it will be important to focus more holistically across the networking stack. Future researchers need to focus more on cross-layer optimization and to study the implications of subtle interplay between various layers.

\subsection{Challenges in modeling CRNs}

%\emph{Challenges in operating in non-Markovian or partially observed environments:}
While it is quite common to use simplistic assumptions (such as the Markovian assumption or the perfect knowledge assumption) to keep our models tractable, real systems are in fact quite complex and CRs often have to work in unknown RF environments. As Benoit Mandelbrot famously lamented `the world unfortunately has not been designed for the convenience of mathematicians', there is a lot of scope of new research in areas of decision making and learning in non-Markovian, partially observed, or unknown environments. In such unknown environments, the usage of model-free methods would become increasingly important.

%\subsection{Operating in multi-agent environments}
%\emph{Challenges in operating in multi-agent environments:}
The cognitive radio networking environment is naturally amenable to distributed \emph{multi-agent} decision making rather than centrally controlled optimization. We have seen earlier how multi-agent environments are much more challenging to design than their single-agent counterparts. Ideas from game-theory and economic market design will become increasingly important as multi-agent learning becomes commonplace in CRN design. With AI-based cognitive networks becoming mainstream, it will be  important to understand the behavior of the overall CRN system in terms of equilibria and dynamics for large distributed networks with multiple learning CRs, each taking self-serving decisions with access to limited information.

\subsection{Eliciting and modeling cooperative behavior}

%\emph{Eliciting and modeling cooperative behavior:}
To effectively perform distributed learning and decision making in wireless networks, cooperative behavior is very important, even for agents interested in personal utility maximization \cite{fitzek2006cooperation}. Already, there has been a lot of work on cooperative spectrum sensing and decision making through coalitions \cite{saad2012cooperative}. In a recently proposed cooperative paradigm, named `docitive networks, it is proposed that agents will learn more efficiently through enhanced cooperative knowledge transfer \cite{giupponi2010docitive}. Docitive networks draw their etymology from the Latin root word \emph{docere} meaning `to teach'. In \cite{giupponi2010docitive}, three distinct docitive approaches were proposed for CRNs: \emph{i)} startup docition, \emph{ii)} adaptive docition, \emph{iii)} iterative docition. In future work, eliciting and encouraging cooperative behavior through incentives and mechanism design will become important and looks promising to be an important area of further research.

\subsection{Understanding the dynamics of CRNs}

Emergent behavior of CRNs, composed of multiple self-interested CR servicing users with distinct context, can be \emph{complex}. This can manifest itself when slight changes in one or more of the system parameters result in dramatic changes in system behavior \cite{haykin2005cognitive}. Researchers can exploit advances in the study of complexity to understand the dynamics of such CRNs \cite{waldrop1992complexity}. The interplay between cooperation, competition, and exploitation has also been explored in \cite{haykin2005cognitive} in which etiquettes and protocols to manage the tradeoff were emphasized.

\vspace{10pt}
\section{\uppercase{Conclusion}}
\label{sec:concl}

Learning is at the core of the vision of cognitive radio and cognitive radio networks. While a lot of previous research attention has focused on general AI techniques for optimizing PHY and MAC layer parameters for CRs, scant attention has been given to utilizing learning techniques at the network layer particularly for the problem of routing. We have argued in this paper that incorporating learning from the past and present conditions can be very productive and can lead to improved CRN performance. In this paper, we have surveyed the set of techniques that can be used to embed learning in the routing framework of CRNs, and provided a tutorial on the various relevant techniques from a wide variety of fields. Open research issues and potential directions for future work have also been identified.

\bibliographystyle{ieeetr}
\bibliography{mlrouting}

\begin{IEEEbiography}[{\includegraphics[width=1in,height=1.25in,clip,keepaspectratio]{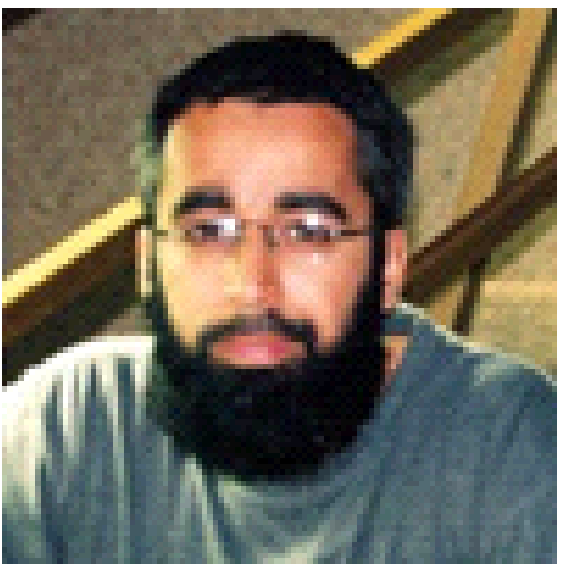}}]{Junaid Qadir}
He is an Assistant Professor in the Electrical Engineering department of the School of Electrical Engineering and Computer Sciences (SEECS), National University of Sciences and Technology (NUST), Pakistan. He is also the Lab Director of the Cognet Lab at SEECS. He completed his BS in Electrical Engineering from UET, Lahore, Pakistan and his PhD from University of New South Wales, Australia. His research interests include networking/ algorithmic issues in cognitive radio networks, wireless networks, and software-defined networks.
\end{IEEEbiography}

\end{document}